%% file: ReorderData-TVCG-main.tex
\documentclass[journal]{IEEEtran}
\usepackage{amsmath,amsfonts}
\usepackage{algorithmic}
\usepackage{algorithm}
\usepackage{array}
\usepackage[caption=false,font=normalsize,labelfont=sf,textfont=sf]{subfig}
\usepackage{textcomp}
\usepackage{stfloats}
\usepackage{url}
\usepackage{verbatim}
\usepackage{graphicx}
\usepackage{cite}
\usepackage{enumerate}
\usepackage{multirow}
\usepackage{color}
\usepackage{lipsum}                    
\usepackage{amssymb,amsmath}
\usepackage{wrapfig}
\usepackage{booktabs}
\usepackage{makecell}
\usepackage{bm}
\usepackage{array}
\usepackage{mathptmx}                  
\usepackage{hyperref}
\usepackage[inline]{enumitem}
\usepackage{pifont}

\def \etal {{\emph{et al}.\thinspace}}

\newcommand{\jiangning}[1]{\textcolor{black}{#1}}
\newcommand{\jiangningminor}[1]{\textcolor{black}{#1}}

\newcommand{\PLH}{{\mkern-2mu\times\mkern-2mu}}
\usepackage{tikz}

\begin{document}

\title{ReorderBench: A Benchmark for Matrix Reordering}

\author{Jiangning Zhu, Zheng Wang, Zhiyang Shen, Lai Wei, Fengyuan Tian, Mengchen Liu, Shixia Liu
}
\markboth{Journal of \LaTeX\ Class Files,~Vol.~14, No.~8, August~2021}%
{Shell \MakeLowercase{\textit{et al.}}: A Sample Article Using IEEEtran.cls for IEEE Journals}


\maketitle

\input{0-abstract}
\input{1-introduction.tex}
\input{2-related.tex}

\input{3-pattern}

\input{4-dataset-construction.tex}

\input{5-dataset-statistics}

\input{6-use-cases.tex}
\input{7-discussion.tex}

\input{8-conclusion.tex}

\section*{Acknowledgments}
This work was supported by the National Natural Science Foundation of China under grant U21A20469 and in part by the Tsinghua University-China Telecom Wanwei Joint Research Center.
The authors would like to thank Duan Li, Zhen Li, and Yukai Guo for their valuable discussions and comments.

\bibliographystyle{IEEEtran}
\bibliography{reference}

\begin{IEEEbiography}
[{\includegraphics[width=1in,height=1.25in,clip,keepaspectratio]{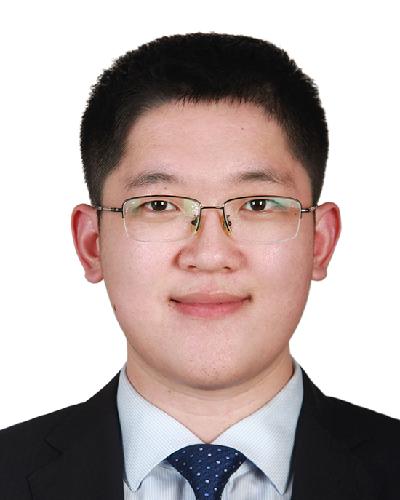}}]
{{Jiangning Zhu}} 
is a second-year Ph.D. student at the School of Software, Tsinghua University. His research interest is explainable artificial intelligence. He received a B.S. degree from Tsinghua University.
\end{IEEEbiography}

\begin{IEEEbiography}
[{\includegraphics[width=1in,height=1.25in,clip,keepaspectratio]{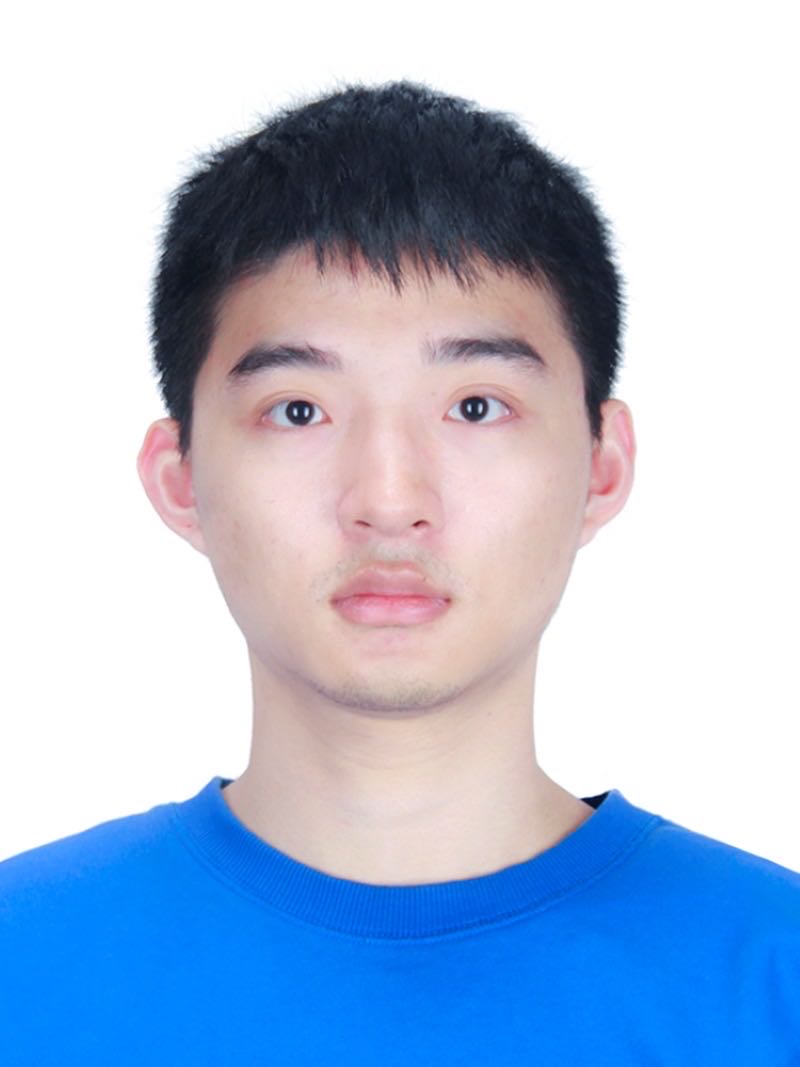}}]
{{Zheng Wang}} is a first-year master's student at the School of Software, Tsinghua University. His research interest is AI for visualization. He received a B.S. degree from Tsinghua University.
\end{IEEEbiography}

\begin{IEEEbiography}
[{\includegraphics[width=1in,height=1.25in,clip,keepaspectratio]{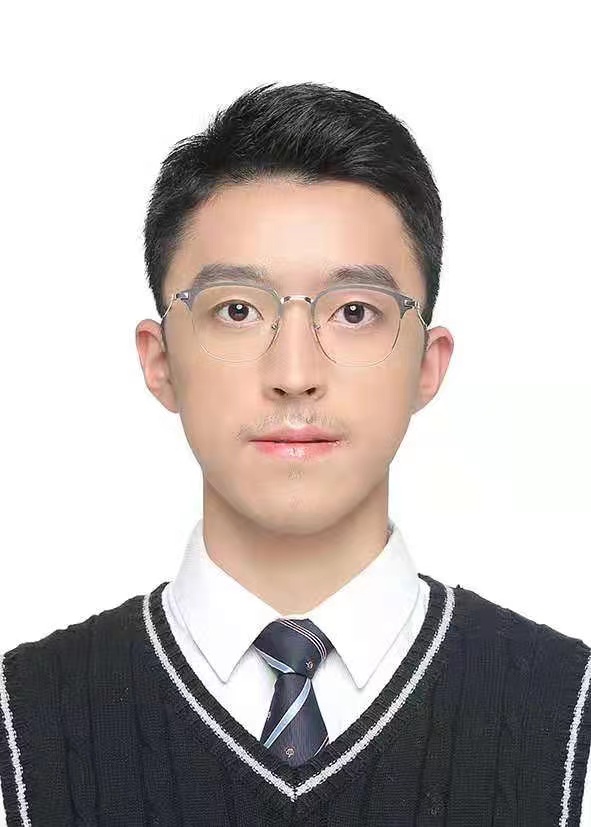}}]{{Zhiyang Shen}} 
is an undergraduate student at the School of Software, Tsinghua University. His research interests include explainable machine learning and visualization.
\end{IEEEbiography}

\begin{IEEEbiography}
[{\includegraphics[width=1in,height=1.25in,clip,keepaspectratio]{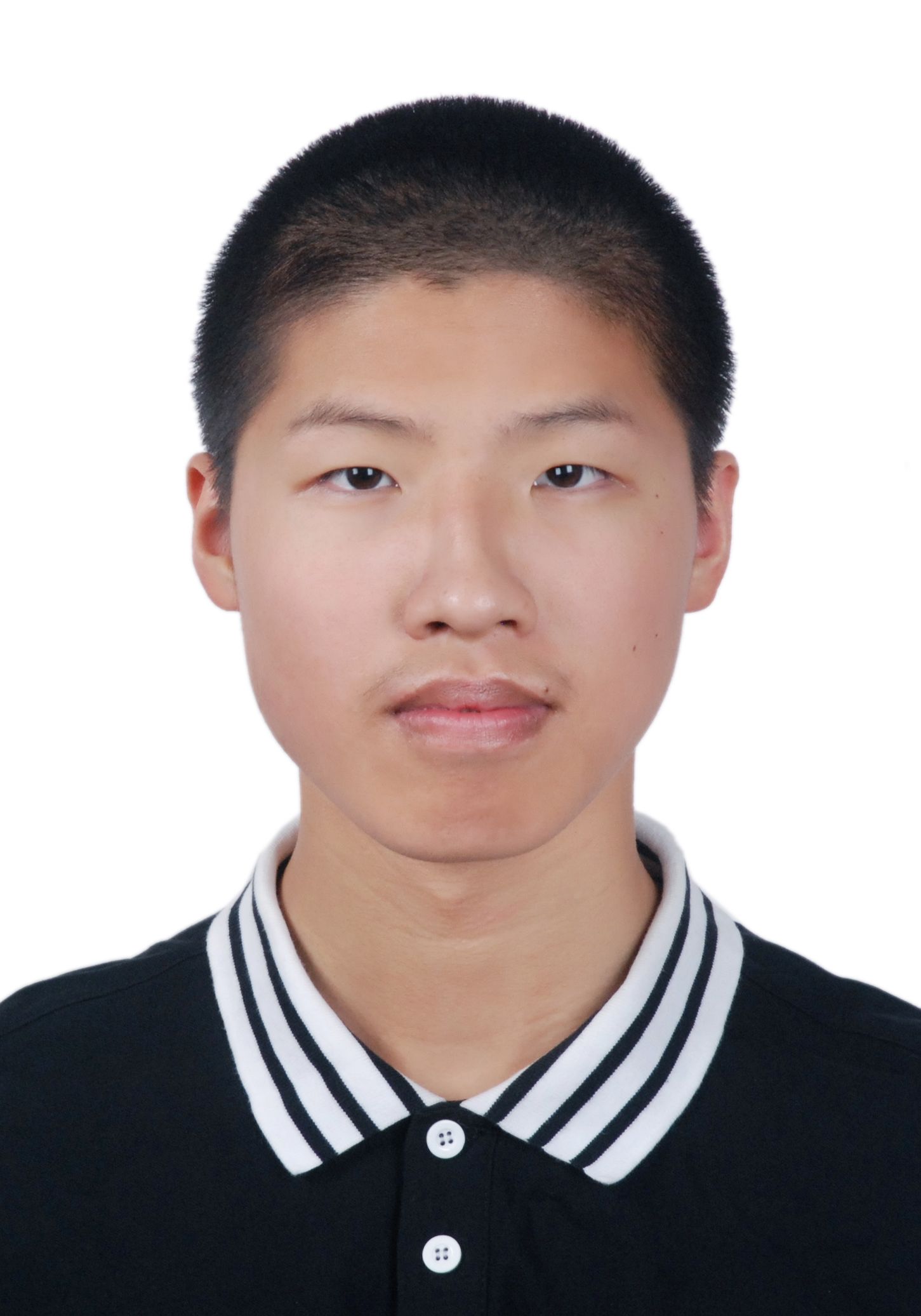}}]
{{Lai Wei}} 
is an undergraduate student at the School of Software, Tsinghua University. His research interests include explainable machine learning, multimodal learning and visualization.
\end{IEEEbiography}

\begin{IEEEbiography}
[{\includegraphics[width=1in,height=1.25in,clip,keepaspectratio]{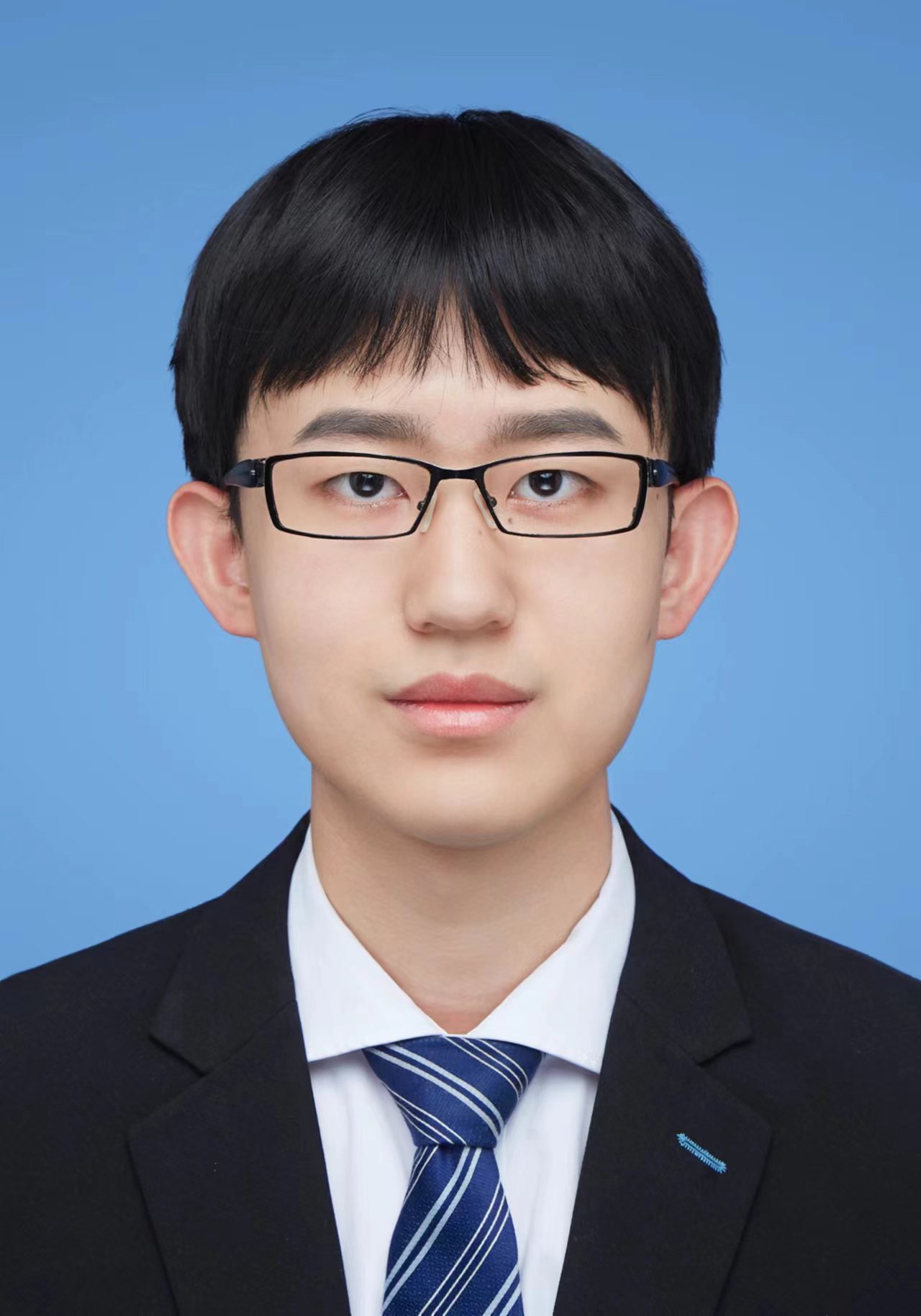}}]
{{Fengyuan Tian}} is a third-year master's student at the School of Software, Tsinghua University.  His research interest is explainable machine learning.
He received a B.S. degree from Tsinghua University.
\end{IEEEbiography}

\begin{IEEEbiography}
[{\includegraphics[width=1in,height=1.25in,clip,keepaspectratio]{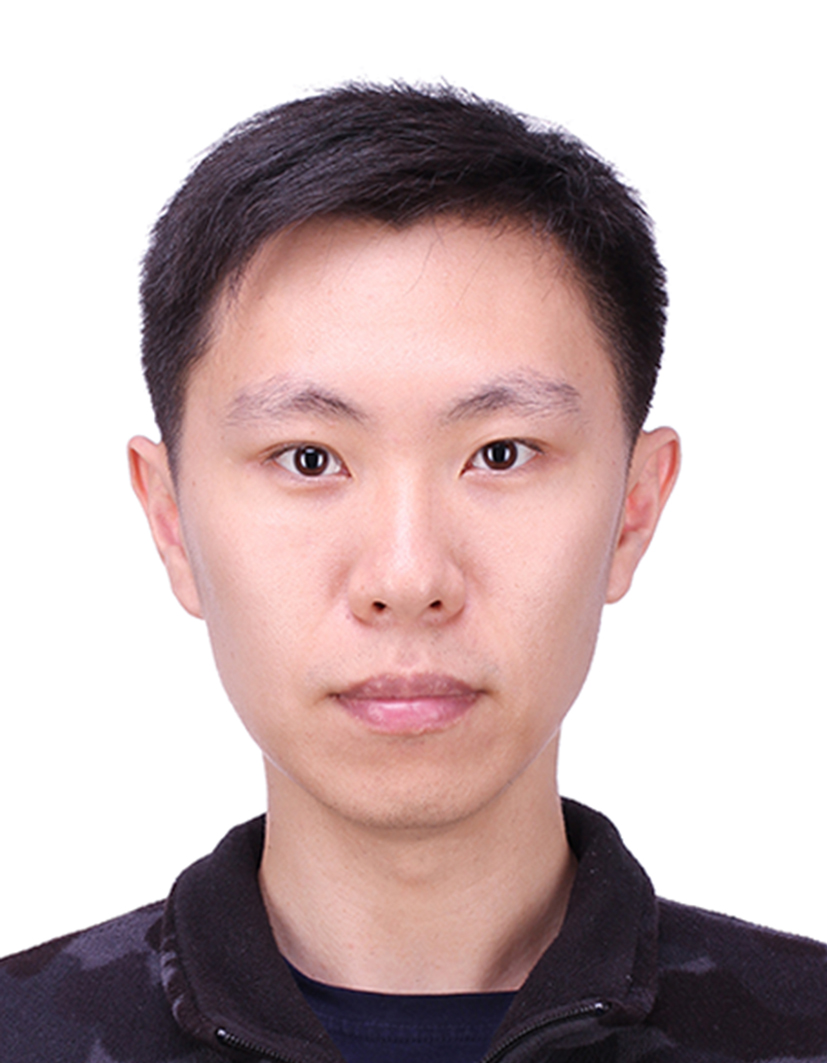}}]{Mengchen Liu} 
is a Senior Researcher at Microsoft. His research interests include explainable AI and computer vision. He received a B.S. in Electronics Engineering and a Ph.D. in Computer Science from Tsinghua University.
He has served as a PC member and reviewer for various conferences and journals.
\end{IEEEbiography}

\begin{IEEEbiography}
[{\includegraphics[width=1in,height=1.25in,clip,keepaspectratio]{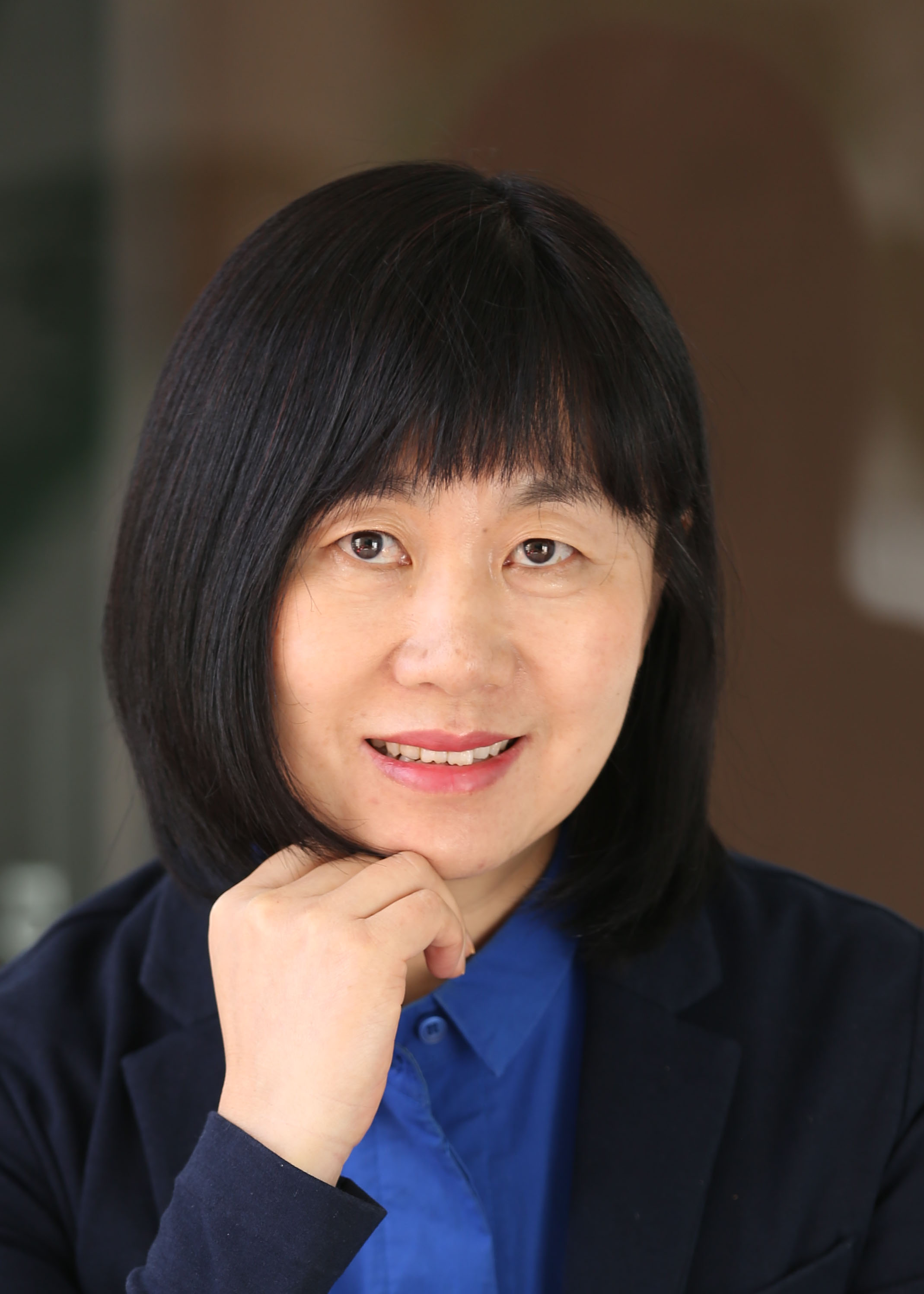}}]{Shixia Liu}
is a professor at Tsinghua University. Her research interests include explainable artificial intelligence, visual analytics for big data. She worked as a research staff member at IBM China Research Lab and a lead researcher at Microsoft Research Asia.
She received a B.S. and M.S. from Harbin Institute of Technology, a Ph.D. from Tsinghua University.
She is a fellow of IEEE and an associate editor-in-chief of IEEE Trans. Vis. Comput. Graph.
\end{IEEEbiography}

\vfill

\end{document}

%% file: 0-abstract.tex
\begin{abstract}
Matrix reordering permutes the rows and columns of a matrix to reveal meaningful visual patterns, such as blocks that represent clusters.
A comprehensive collection of matrices, along with a scoring method for measuring the quality of visual patterns in these matrices, contributes to building a benchmark.
This benchmark is essential for selecting or designing suitable reordering algorithms for \jiangning{revealing specific patterns}.
In this paper, we build a \jiangning{matrix-reordering} benchmark, ReorderBench, with the goal of evaluating and improving \jiangning{matrix-reordering} techniques.
This is achieved by generating a large set of representative and diverse matrices and scoring these matrices with a convolution- and entropy-based method. 
Our benchmark contains 2,835,000 binary matrices and 5,670,000 continuous matrices, each \jiangning{generated to exhibit} one of four visual patterns: block, off-diagonal block, star, or band, \jiangning{along with 450 real-world
matrices featuring hybrid visual patterns}.   
We demonstrate the usefulness of ReorderBench through three main applications in matrix reordering: 1) evaluating different reordering algorithms, 2) creating a unified scoring model to measure the visual patterns in any matrix, and 3) developing a deep learning model for matrix reordering.
\end{abstract}

\begin{IEEEkeywords}
Matrix reordering, visual pattern, benchmark, binary matrix, continuous matrix
\end{IEEEkeywords}

%% file: 1-introduction.tex
\section{Introduction}
\maketitle
Matrix reordering permutes the rows and columns of a matrix to reveal meaningful visual patterns, such as blocks for clusters, off-diagonal blocks for bi-cliques, stars for highly connected nodes, and bands for paths.
Despite its significant potential to enhance data visualization and analysis, most existing algorithms are limited by being developed \jiangning{for understandable metrics} \cite{bar2001fast, earle2015advances, watanabe2022deep} or large sets of matrices that share the same visual pattern \cite{kwon2022deep}. 
Due to the lack of extensive and diverse matrices with an appropriate scoring method to measure the quality of their visual patterns makes it difficult to develop an effective reordering method for revealing various patterns.
Although Behrisch~\etal~\cite{behrisch2016magnostics} have taken the first step in building a \jiangning{matrix-reordering} dataset, this dataset has two limitations.
First, their focus has been limited to binary matrices with entries of $0$ or $1$, while real-world applications often involve continuous matrices with real-valued entries. 
Second, the scoring method \jiangning{in their dataset} is \jiangning{based on the level of degeneration introduced during generation, such as the number of index swaps in the index-swap function. 
This scoring method is intrinsically} tied to the \jiangning{matrix-generation} process and cannot directly measure the quality of visual patterns in matrices reordered by different algorithms.

To overcome these limitations, it is essential to build a \jiangning{matrix-reordering} benchmark that includes both binary and continuous matrices, along with a more generic scoring method capable of evaluating the quality of different visual patterns in matrices reordered by different algorithms.
Such a benchmark will facilitate the evaluation and development of \jiangning{matrix-reordering} techniques.
First, it enables the selection of suitable reordering algorithms for \jiangning{revealing the underlying visual patterns} by evaluating their reordering performance.
Second, it supports the development of new reordering methods that work well for different visual patterns.

To better support these tasks, we build the ReorderBench benchmark
by generating a large collection of representative and diverse symmetric matrices and scoring them with a convolution- and entropy-based method. 
To ensure the representativeness and diversity of the benchmark, we first generate a set of representative matrix templates for each visual pattern.
Then, based on these matrix templates, a large number of matrix variations with diverse degrees of degeneration are generated. 
The diversity is achieved by combining different variation methods, including noise addition and index swapping. 
To accurately evaluate the quality of visual patterns in a matrix, we develop a scoring method by combining the matching capability of convolutional kernels~\cite{lecun1998gradient} and the disorder detection capability of entropy~\cite{bishop2006pattern}. 
ReorderBench contains $2,835,000$ binary matrices and $5,670,000$ continuous matrices, \jiangning{each generated to exhibit one of} four patterns: block, off-diagonal block, star, or band, \jiangning{along with $450$ real-world matrices featuring hybrid visual patterns.}
Table~\ref{table:comparisson} provides a statistical comparison with the existing matrix datasets.
\jiangning{A more detailed version of the comparison is provided in Table 1 of the supplemental material.}

\begin{table}[!t]
\caption{A comparison with the existing datasets. "Quality score?" refers to whether the dataset contains quality scores. "\jiangning{Direct use?}" refers to whether \jiangning{its quality score} can be directly used to measure the quality of reordered matrices.}
\vspace{-2mm}%
\fontsize{8}{8}\selectfont
\centering
\renewcommand\arraystretch{1.1}
\setlength{\tabcolsep}{.3em}{
\begin{tabular}{ccccc}
\toprule
Dataset & \#\ Binary & \#\ Continuous & Qualtiy score? & \jiangning{Direct use?}\\ \midrule
ReorderBench & \jiangning{$2,835,450$} & $5,670,000$ & \checkmark & \checkmark \\
\begin{tabular}[c]{@{}c@{}}Magnostics~\cite{behrisch2016magnostics} \end{tabular} & $5,570$ & $0$ & \checkmark & $-$ \\ 
Pajek Graph~\cite{batagelj1998pajek} & $44$ & $32$ & $-$ & $-$ \\
Petit Testsuite~\cite{petit2003experiments} & $21$ & $0$ & $-$ & $-$ \\
Matrix Market~\cite{boisvert1997matrix} & $112$ & $386$ & $-$ & $-$ \\
SuiteSparse~\cite{davis2011university} & $601$ & $2,292$ & $-$ & $-$ \\
Network Repo.~\cite{rossi2015network} & $3,624$ & $3,030$ & $-$ & $-$ \\
\bottomrule
\end{tabular}}
\vspace{-4mm}
\label{table:comparisson}
\end{table}

We demonstrate the usefulness of ReorderBench through three applications.
First, we evaluate the capability of different reordering algorithms to reveal visual patterns using the ReorderBench test set. 
Second, we develop a unified scoring model to measure the quality of visual patterns in any matrix, including matrices beyond ReorderBench. 
Third, we build a deep learning model for matrix reordering to better reveal the inherent visual patterns in matrices.

The main contributions of this work are threefold:
\begin{itemize}
    \item A pipeline for generating a representative and diverse collection of binary and continuous matrices along with the quality scores of their visual patterns.
    \item A \jiangning{matrix-reordering} benchmark for selecting or designing an appropriate reordering algorithm for \jiangning{revealing the underlying visual patterns}, available at: \href{https://reorderbench.github.io/}{https://reorderbench.github.io/}.
    \item Three applications for demonstrating the usefulness of our benchmark in evaluating reordering algorithms, creating a unified scoring model, and developing a deep model for matrix reordering.
\end{itemize}

%% file: 2-related.tex
\section{Related Work}
\label{sec:related-work}
Our work is related to \jiangning{matrix-reordering} methods, quality metrics, and matrix dataset.

\vspace{1mm}
\noindent \textbf{\jiangning{Matrix-reordering} method}.
Matrix reordering is a long-standing research problem that has received attention from multiple fields, such as sociology and bioinformatics~\cite{liiv2010seriation}.
Earlier efforts focus on developing classical algorithms to optimize a given quality metric \jiangning{for an individual matrix}.
According to the optimization strategy, these algorithms can be categorized into two classes: exact algorithms~\cite{brusco2005optimal,brusco2008heuristic} and approximate algorithms~\cite{gruvaeus1972two,earle2015advances,morris2003dendrogram,friendly2002corrgrams,rodgers1992seriation}.
Exact algorithms include the branch-and-bound algorithm~\cite{brusco2005optimal} and the dynamic programming algorithm~\cite{brusco2008heuristic}.
However, because of their exponential time complexity, obtaining exact solutions for matrix reordering becomes infeasible as the matrix size increases.
Therefore, approximate algorithms based on various heuristics have been proposed.
To place similar rows and columns close to each other, Gruvaeus~\etal~\cite{gruvaeus1972two} reorder matrices through hierarchical clustering.
Later studies focus on refining the resulting hierarchical clustering dendrogram using a greedy strategy~\cite{earle2015advances} or simulated annealing~\cite{morris2003dendrogram}.
Another line of work utilizes dimension reduction techniques, such as principal component analysis~\cite{friendly2002corrgrams} and multi-dimensional scaling~\cite{rodgers1992seriation}, to capture the data similarity in a lower-dimensional space.
The rows are projected onto one dimension, and the matrix is reordered accordingly.
For more details on classical algorithms \jiangning{reordering individual matrices}, the readers are referred to the survey by Behrisch~\etal~\cite{behrisch2016matrix}.
\jiangning{Another line of work focuses on reordering collections of matrices~\cite{van2021simultaneous,van2024contextual}.
For example, Beusekom \etal~\cite{van2024contextual} aim to achieve contextual orderings that balance the consistency across the collection and the quality of individual reordering results.}

Recently, to avoid the trial-and-error process of choosing an appropriate algorithm for the unknown visual patterns in a matrix, several deep learning frameworks have been developed.
Building upon dimension-reduction-based reordering methods, Watanabe~\etal~\cite{watanabe2022deep} use an autoencoder to encode the rows and columns of a given matrix into one-dimensional features.
The rows and columns are then reordered in ascending order of the one-dimensional features.
Kwon~\etal~\cite{kwon2022deep} introduce the use of a variational autoencoder to generate different reordering results for a matrix. 
However, due to the lack of reordering benchmarks with diverse matrices, existing deep learning models are typically trained \jiangning{on limited datasets that consist of either the given matrix~\cite{watanabe2022deep} or its reordering results~\cite{kwon2022deep}.
These training methods aim to learn features specific to the given matrix, resulting in models that are adept at reordering only the given matrix and thus lack generalizability}.
ReorderBench addresses this issue by providing representative and diverse matrices along with accurate quality scores.
Upon this benchmark, we build a \jiangning{matrix-reordering} model that is generalized to previously unseen matrices.

\vspace{1mm}
\noindent \textbf{Quality metric}.
Quality metrics measure the quality of visual patterns in matrices and serve as the optimization criteria for reordering algorithms. 
Most existing quality metrics implement the idea of placing similar rows and columns close to each other to reveal visual patterns~\cite{behrisch2019guiro}.
Generally, they fall into two categories: adjacency-based metrics~\cite{mccormick1972problem,van2021simultaneous} and \jiangning{distance}-based metrics~\cite{diaz2002survey,harper1966optimal,petit2003experiments,robinson1951method,chen2002generalized,hubert2001combinatorial,hubert1976quadratic,earle2015advances}.

Adjacency-based metrics \jiangning{measure the similarity between adjacent entries in the matrix}.
An example is the measure of effectiveness (ME)~\cite{mccormick1972problem}, which is calculated as the sum of the product of adjacent entries.
In matrices with high measure of effectiveness, the entries are clustered to form block or off-diagonal block patterns.
\jiangning{More recently, Moran's \textit{I}~\cite{van2021simultaneous} has been applied to measure the spatial auto-correlation in matrix entries.
It classifies adjacencies between entries with values of 0 and 1 into three categories: $1$-$1$, $0$-$0$, and $0$-$1$, and calculates a weighted sum of their occurrences. 
To accommodate matrices with varying levels of sparsity, the weights are determined based on the sum of all entries.
}

\jiangning{Distance}-based metrics \jiangning{focus on the distances between the rows and columns in the ordering}.
\jiangning{Some of these metrics measure how well the ordering distances align with the connectedness in the underlying graph~\cite{diaz2002survey}.
For example, the linear arrangement (LA)~\cite{petit2003experiments} measures the sum of the distances between the connected vertices in the underlying graph}.
\jiangning{Other distance-based metrics measure how well the ordering distances align with the dissimilarity between rows and columns based on the dissimilarity matrix.}
Robinson's seminal work~\cite{robinson1951method} defines a perfectly ordered dissimilarity matrix as an anti-Robinson matrix, where dissimilarity values monotonically increase away from the main diagonal in all rows and columns.
Following this, many metrics have been developed to assess deviations from the anti-Robinson matrix, including anti-Robinson events/deviation (AR events/deviation)~\cite{chen2002generalized} and gradient measures~\cite{hubert2001combinatorial}.
Parallel to these metrics, several \jiangning{distance}-based metrics directly measure the distance of the large dissimilarity values from the main diagonal.
For example, the linear seriation criterion~\cite{hubert1976quadratic} is defined as the sum of the product of the values and their distance from the main diagonal.
Recently, Earle~\etal~\cite{earle2015advances} propose the banded anti-Robinson form (BAR) to better measure the quality of local visual patterns.
This metric is a relaxed version of the linear seriation criterion.

These adjacency- or \jiangning{distance}-based metrics work well for block and off-diagonal block patterns.
However, \jiangning{as they typically focus on placing similar rows and columns close to
each other}, they are less effective in measuring other patterns\jiangning{, such as star patterns and band patterns}, \jiangningminor{where even small amounts of noise can notably distort row and column similarities}.
To ensure that more visual patterns are better revealed, an effective metric is required. 
To this end, we propose the convolution- and entropy-based scoring method, \jiangning{which combines the matching capability of convolutional kernels~\cite{lecun1998gradient} and the disorder detection capability of entropy~\cite{bishop2006pattern}} to directly measure the quality of visual patterns in ReorderBench.
\jiangning{This scoring method accurately evaluates the quality of visual patterns in a matrix, and is essential for constructing a benchmark that supports the evaluation and development of matrix-reordering techniques.}
We also develop a unified scoring model to measure visual patterns in any matrix, including matrices beyond ReorderBench.

\vspace{1mm}
\noindent \textbf{Matrix dataset}.
Many matrix datasets are publicly available, including the Pajek graph collection~\cite{batagelj1998pajek}, the Petit Testsuite~\cite{petit2003experiments}, Matrix Market~\cite{boisvert1997matrix}, the SuiteSparse matrix collection~\cite{davis2011university}, and the Network Repository~\cite{rossi2015network}.
Existing \jiangning{matrix-reordering} methods often use some matrices from these datasets to showcase their reordering capability.  
However, the lack of quality metrics in these datasets prevents a quantitative evaluation of reordering methods.
Consequently, there is a need for a benchmark that includes both matrices and quality metrics.

Behrisch~\etal~\cite{behrisch2016magnostics} have pioneered the creation of the Magnostics dataset, which consists of $5,570$ binary matrices.
It aims to evaluate the capability of hand-crafted features in detecting four visual patterns (block, off-diagonal block, star, and band) and two anti-patterns (bandwidth and noise).
The dataset is created by first generating a type of visual pattern on empty matrices.
The matrices are then gradually degenerated using one of the variation methods, including point-swap, index-swap, and masking.
Quality scores are assigned to the matrices based on the level of degeneration introduced by the function, such as the number of index swaps in the index-swap function.
This dataset works well to select \jiangning{high-dimensional} features capable of detecting visual patterns in matrices \jiangning{using distance-based search}.
Although the selected features could \jiangning{potentially} be used to \jiangning{derive a quality metric for evaluating} reordering algorithms for binary matrices, the authors have not fully explored the exact method.
Furthermore, the dataset only includes binary matrices, which limits their usage in real-world applications.
In contrast, ReorderBench introduces a convolution- and entropy-based scoring method.
Unlike the scoring method used in Magnostics, it does not rely on the variation methods that degenerate the matrices. 
As a result, it can be directly used to compare reordering algorithms.
ReorderBench also extends to generating and scoring continuous matrices, addressing the gap left by the Magnostics dataset.

%% file: 3-pattern.tex
\section{Visual Pattern Summary}
\label{sec:pattern}

There are four commonly used visual patterns in matrices: block, off-diagonal block, star, and band~\cite{behrisch2016matrix,behrisch2019guiro,behrisch2016magnostics,van2021simultaneous}.
\jiangning{Despite the ongoing debate regarding the interpretation of band~\cite{shu2024does}, we include it in our work due to its application in several important fields, such as bioinformatics~\cite{zheng2022normalization} and network analysis~\cite{wozniak2022new}.
This inclusion also serves to demonstrate the generalizability of our matrix generation and scoring methods. 
For applications where the band patterns are not required, they can be excluded from our benchmark.}

\setlength\intextsep{0pt}
\setlength\columnsep{1mm}
\begin{wrapfigure}[4]{l}{0.375\columnwidth}
\includegraphics[width=0.375\columnwidth]{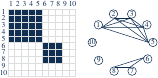}
\end{wrapfigure}

\begin{figure*}[!tb]
  \centering
  \includegraphics[width=\linewidth]{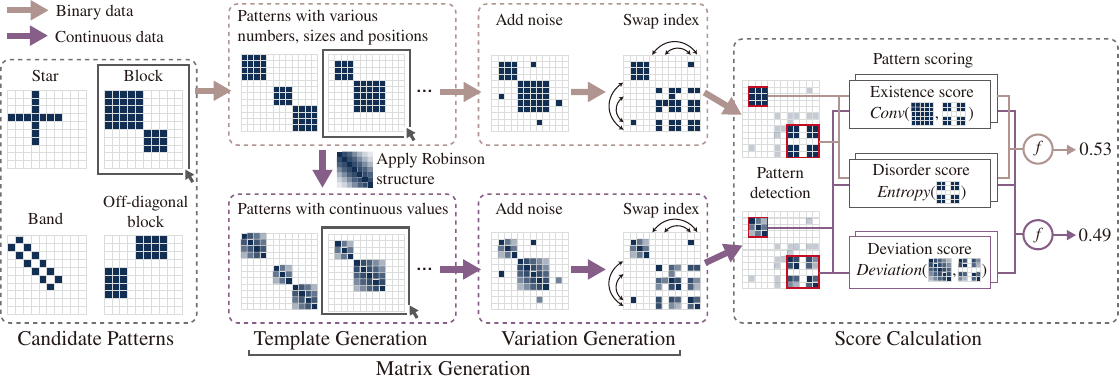}
  \vspace{-3mm}
  \caption{The generation pipeline for the ReorderBench benchmark.}
  \label{fig:system_pipeline}
  \vspace{-5mm}
\end{figure*}

\noindent\textbf{Block pattern}. A block pattern is characterized by a square area situated along the main diagonal of the matrix.
It represents densely connected clusters in the underlying graph, where nodes are connected to each other.
The block pattern can represent groups of mutual friends in social networks or regions with similar pixel values in images.

\setlength\intextsep{0pt}
\begin{wrapfigure}[4]{l}{0.375\columnwidth}
\includegraphics[width=0.375\columnwidth]{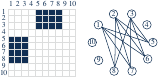}
\end{wrapfigure}

\noindent\textbf{Off-diagonal block pattern}. 
Similarly to the block pattern, an off-diagonal block pattern is a rectangular area that does not touch the main diagonal.
It indicates that there exist bi-cliques in the underlying graph.
Nodes in a bi-clique are divided into two disjoint sets, where each node in one set is connected to each node in the other set. 
For example, in ecology, this pattern appears in species-interaction networks, where one set represents predator species, and the other set represents prey species. 
The edges between two sets indicate predation interactions.

\setlength\intextsep{0pt}
\begin{wrapfigure}[4]{l}{0.375\columnwidth}
\includegraphics[width=0.375\columnwidth]{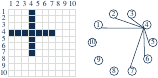}
\end{wrapfigure}
\noindent\textbf{Star pattern}. A star pattern is characterized by two intersecting lines, one horizontal and one vertical.
Either of these lines does not need to span the whole matrix.
A star pattern represents a node with many connections in the graph.
Typical examples include a highly influential individual who is connected to many other individuals in a social network and a data center connected to multiple client devices in an internet network. 

\setlength\intextsep{0pt}
\begin{wrapfigure}[4]{l}{0.375\columnwidth}
\includegraphics[width=0.375\columnwidth]{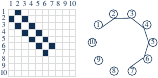}
\end{wrapfigure}
\noindent\textbf{Band pattern}. A band pattern is characterized by lines that run parallel to the main diagonal of the matrix.
It indicates \jiangning{the presence of paths, cycles, or meshes} in the underlying graph.
For example, in a social network, a path can illustrate the spread of information, opinions, or behaviors through a sequence of links from one node to another.

%% file: 4-dataset-construction.tex
\section{Benchmark Generation}
\label{sec:construction}
ReorderBench contains both binary and continuous matrices.
Fig.~\ref{fig:system_pipeline} shows its generation pipeline,
which includes two main steps: matrix generation and score calculation.

\subsection{Matrix Generation}
\label{subsec:generation}
This step aims to generate a collection of representative and diverse matrices \jiangningminor{with non-overlapping patterns}. 
\jiangning{Since there is a one-to-one correspondence between \jiangningminor{these} matrices and their underlying graphs, this process also ensures the representativeness and diversity of graph structures.}
We achieve representativeness by generating various matrix templates for each visual pattern (\textbf{template generation}).
We ensure diversity by degenerating each template into a set of matrix variations through methods such as adding noise and swapping indices (\textbf{variation generation}).

\subsubsection{Template Generation}
\begin{figure}[!tb]
\centering
{\includegraphics[width=\linewidth]{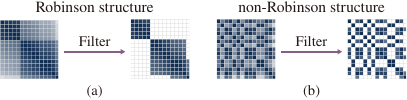}}
\vspace{-3mm}
\caption{The comparison of the structures after filtering: (a) the Robinson structure highlights underlying visual patterns; (b) the non-Robinson structure shows degenerated underlying visual patterns.}
\vspace{-5mm}
\label{fig:filtering}
\end{figure}

Binary and continuous matrices exhibit similar visual patterns introduced in Sec.~\ref{sec:pattern}.
Thus, the process of generating templates for both types can be largely unified. 
Previous research on continuous matrices has revealed the distinctiveness of their visual patterns, notably characterized by the Robinson structure~\cite{tien2008methods}. 
In this structure, entries in a block or star pattern monotonically decrease away from the main diagonal.
An off-diagonal block pattern is regarded as a mirrored version of a block pattern, with its larger entries closer to a diagonal that mirrors the main diagonal of the block pattern.
For a band pattern, the Robinson structure is not applicable, as its entries are parallel to the main diagonal.
This structural property exhibits two advantages in the analysis of matrices.
First, it illustrates a key characteristic of fully reordered continuous matrices: similar rows are closely positioned~\cite{robinson1951method}.
This aligns with human perception and allows users to better identify visual patterns~\cite{van2021simultaneous}.
Moreover, a matrix featuring this structure is representative of those with non-Robinson structures whose optimal reordering results are the same as this matrix. 
Second, it facilitates a widely used analytical technique for continuous matrices: filtering low-valued entries to highlight underlying visual patterns~\cite{wu2010gap}.
As shown in Fig.~\ref{fig:filtering}, the Robinson structure preserves the underlying visual pattern more effectively than the non-Robinson structure when low-valued entries are filtered. 
A straightforward method for generating continuous templates is to replace the $1$s in a binary template with random values in $[0,1]$.
However, this method often destroys the Robinson structures, limiting the representativeness of the resulting templates by excluding those based on Robinson structures.
\jiangning{Previous works have shown that starting with representative examples and diversifying them is a practical method for constructing effective datasets~\cite{tripathi2019learning,shivashankar2023semantic}. 
Thus, to obtain a set of representative continuous matrix templates}, we first generate a large set of binary templates and then apply the Robinson structure to produce various continuous ones.
This method effectively preserves the representativeness of the templates by including all fully reordered continuous matrices derived from the binary templates.

\begin{figure}[!b]
\centering
\vspace{-3mm}
{\includegraphics[width=\linewidth]{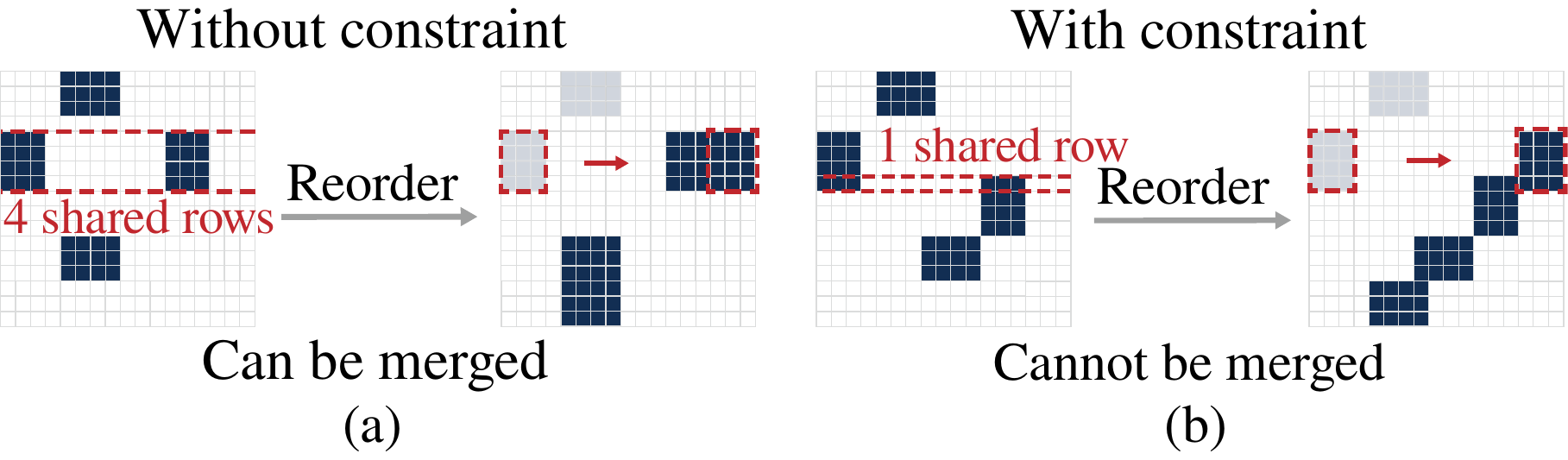}}
\vspace{-3mm}
\caption{The comparison of generated matrix templates with and without the position constraint: (a) without the constraint, two off-diagonal blocks can merge into a larger block; (b) with the constraint, such merging is prevented.}
\label{fig:constraint}
\end{figure}

\vspace{1.5mm}
\noindent\textbf{Binary templates}.
The representativeness of the generated binary templates is ensured by aligning their statistics with real-world matrices, including the number and sizes of the visual patterns in each matrix.
In ReorderBench, we focus on generating matrices with one type of visual pattern.
As shown in Fig.~\ref{fig:system_pipeline}, for each template, we first select the type of visual pattern to appear.
Then, we determine the number of visual patterns in the template.
The number of visual patterns in each template ranges from $1$ to $15$ because we find that in $1,217$ real-world matrices randomly selected from the \jiangning{$6,654$} matrices in the Network Repository~\cite{rossi2015network}, 98.2$\%$ have no more than 15 visual patterns.
\jiangning{According to an established statistical sampling theory~\cite{Krejcie1970determining}, the confidence level of this estimation is given by: $P(\chi^2 \leq 0.01 \cdot n \cdot (N-1) / (N-n))$, where $n=1,217$ is the number of sampled matrices, $N=6,654$ is the total number of matrices, and $\chi^2$ is a chi-square random variable with 1 degree of freedom.
Based on this theory, the proportion estimation of $98.2\%$ has a confidence level of over $99.9\%$, which confirms its reliability.}
Given the number of visual patterns to generate, we randomly select their sizes and positions.
We impose no constraint on the sizes of the visual patterns except for setting the maximum width as 4 for the star and band patterns.
This is derived from our annotations of matrices with a total of ${685}$ star and band patterns, where $99.0\%$ of them satisfy this constraint.
The positions of the patterns are selected to ensure that they do not overlap.
Moreover, for off-diagonal blocks, we ensure that the number of shared rows between two patterns does not exceed half of the number of their rows.
This constraint prevents two off-diagonal blocks from being reordered to form a larger one, thus \jiangning{avoiding unintended patterns and} ensuring the optimal ordering of the templates.
For example, as shown in Fig.~\ref{fig:constraint}(a), two off-diagonal blocks with 4 shared rows can be reordered to form a larger one.
In contrast, Fig. ~\ref{fig:constraint}(b) shows that with only 1 shared row, it is impossible to reorder them into a larger off-diagonal block.
\jiangningminor{To accommodate a broader range of matrices, we do not constrain consistent values on the diagonal. 
For applications focusing on the undirected graph, the diagonal values can be consistently set to 0 or 1, following the common practice.
}

\begin{figure}[!tb]
\centering
{\includegraphics[width=\linewidth]{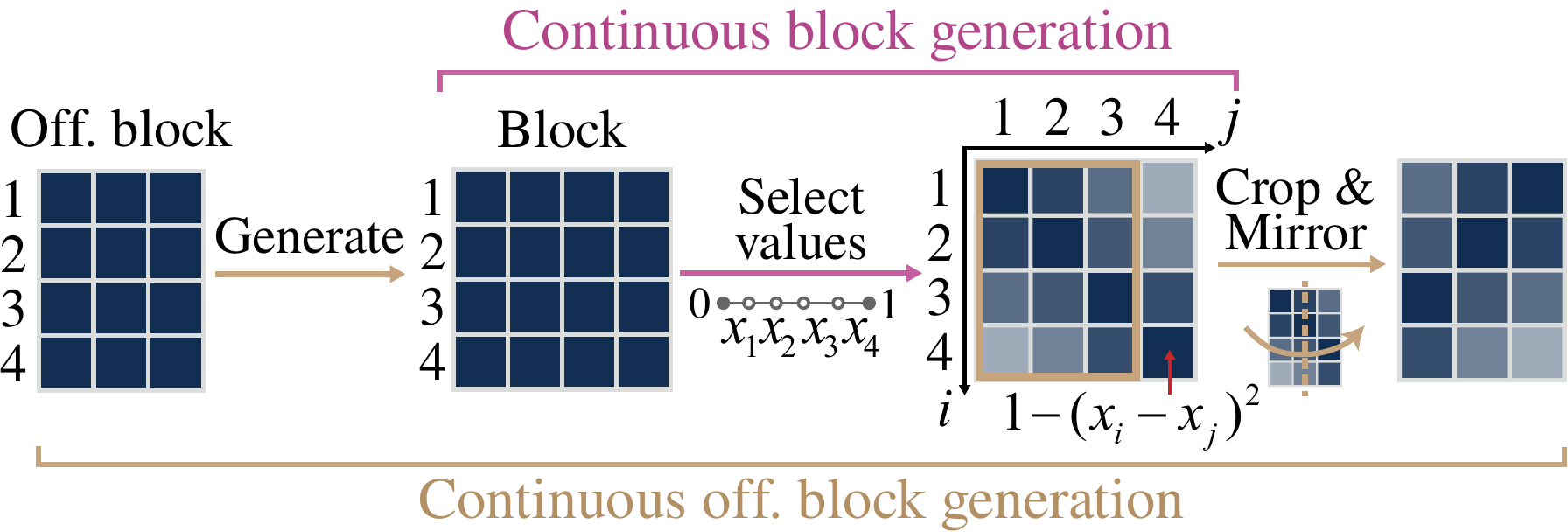}}
\vspace{-3mm}
\caption{The generation of the Robinson structure for visual patterns.}
\label{fig:continuous_pattern}
\vspace{-5mm}
\end{figure}

\vspace{1.5mm}
\noindent\textbf{Continuous templates}.
To generate a continuous template, we first generate a binary one.
Then, as shown in Fig.~\ref{fig:continuous_pattern}, we apply the Robinson structure and replace the $1$s in a binary template with values in $[0,1]$.
The Robinson structure of a block or star pattern is generated by reversing unidimensional scaling, a widely used method in matrix generation~\cite{hahsler2008getting}. 
For a pattern with $u$ rows, we first randomly select a set of ascending values $x_1\leq x_2\leq \dots \leq x_u$ from the range $[0,1]$.
Then, we replace all $1$s in the pattern, where an entry on the $i$-th row and the $j$-th column is replaced by $1-(x_i-x_j)^2$.
For an off-diagonal block pattern with $u$ rows and $v$ columns, we first generate a block pattern with $max(u,v)$ rows.
We then crop and mirror it to obtain the off-diagonal block pattern.
Band patterns are parallel to the main diagonal, 
preventing the application of the Robinson structure. 
Thus, the entries in a band pattern are replaced by random values from $[0,1]$.
\jiangningminor{For applications where values above a certain threshold are considered meaningful, such as those interpreting entries as strength or likelihood of connectivity, the entries can be sampled from a constrained range (\textit{e.g.}, [0.1,1]) to strictly ensure that they exceed a minimum threshold.}

\subsubsection{Variation Generation}
\label{subsubsec:variation}
Previous research has shown that adding noise and swapping indices are effective in increasing the diversity of matrices~\cite{behrisch2016magnostics}.
Consequently, we use these two methods to degenerate each matrix template into a set of matrix variations, thereby enhancing the diversity of the benchmark.
To ensure symmetry, these variations are simultaneously applied to diagonal symmetric entries.

\begin{figure}[!b]
\centering
\vspace{-3mm}
{\includegraphics[width=\linewidth]{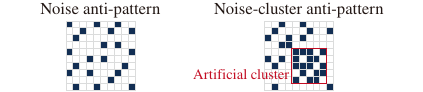}}
\vspace{-3mm}
\caption{The two main intrinsic anti-patterns in matrices.}
\label{fig:antipattern}
\end{figure}

\vspace{1.5mm}
\noindent\textbf{Adding noise}. Behrisch~\etal~\cite{behrisch2019guiro} have found that combining noise in the form of anti-patterns with visual patterns increases the diversity of matrices.
Therefore, we utilize such noise to degenerate matrix templates into a set of variations.
As shown in Fig.~\ref{fig:antipattern}, two main intrinsic anti-patterns exist in the matrices: noise and noise-cluster.
Noise anti-patterns are characterized by the random distribution of non-zero entries throughout the matrix, which lacks inherent visual patterns.
Conversely, noise-cluster anti-patterns form artificial clusters when noise aggregates. 
Each cluster consists of similar yet not identical rows.
\jiangning{The key distinction between the noise-cluster anti-patterns and actual clusters in the matrix template lies in their purpose in the matrix generation process. 
Actual clusters are intended to convey meaningful relationships, whereas noise-cluster anti-patterns serve to degenerate patterns in the template by simulating structured noise or anomalies within the data.
\jiangningminor{Addressing noise-cluster anti-patterns is crucial for improving reordering performance, as they frequently obscure meaningful patterns and challenge existing methods~\cite{behrisch2019guiro}.}
}
Three steps for adding noises are: 
\begin{enumerate*}
\item determining noise levels,
\item noise generation, and
\item noise application. 
\end{enumerate*}

\begin{figure}[!b]
\centering
\vspace{-3mm}
{\includegraphics[width=\linewidth]{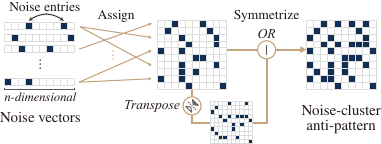}}
\vspace{-3mm}
\caption{The generation of the noise-cluster anti-patterns.}
\label{fig:noise_cluster}
\end{figure}

The upper bound for the noise levels is set to the maximum noise level that can visually preserve the patterns.
Through analysis of matrices subjected to noises of various levels, we find that $16\%$ is an appropriate upper bound. 
The detailed analysis is provided in \jiangning{Sec. 2 of} the supplemental material.
Uniformly distributed noise levels are commonly used in 
dataset generation~\cite{paulin2023review}.
Therefore, we use a set of uniform noise levels:
$[0\%,1\%,\dots,16\%]$.
For each matrix variation, we randomly select two noise levels from this set, one for each type of anti-pattern.
For each matrix template, a sufficient number of variations should be generated to cover the set of noise levels.
We derive in \jiangning{Sec. 3 of} the supplemental material that approximately $70$ variations are required in our case.

Based on the selected noise levels, we generate noise anti-patterns and noise-cluster anti-patterns separately.
For noise anti-patterns, we randomly introduce non-zero entries throughout the matrix according to the noise level.
For noise-cluster anti-patterns, a straightforward method is to generate several clusters of similar yet not identical rows and then apply index swaps to simultaneously distribute the cluster-associated rows throughout the matrix.
However, in such anti-patterns, all artificial clusters can be recovered simultaneously by reversing the index swaps.
They will severely distort the visual patterns in the template.
To tackle this issue, we generate and distribute each cluster separately so that they will be recovered by different orderings of the matrix rows and columns.
Specifically, as shown in Fig.~\ref{fig:noise_cluster}, we first generate a set of $n$-dimensional noise vectors, where $n$ is the number of rows in the matrices.
Each noise vector randomly contains several non-zero entries determined by the specified noise level.
Then, for each row of the anti-pattern, we randomly select one of these noise vectors.
The rows with the same noise vector form a cluster. 
Finally, we symmetrize the anti-pattern by performing a logical OR with its transposed version. 
This symmetrization process ensures that rows within the same cluster are similar yet not identical.
To moderate the impact of noise-cluster anti-patterns on visual patterns across matrix templates \jiangningminor{and ensure they do not unintentionally form meaningful visual patterns}, choosing an appropriate set size for the noise vectors is crucial.
If the set is too small, a single noise vector can appear in too many rows.
As a result, the rows with this noise vector form large clusters, which severely distort the visual patterns in the template. 
On the other hand, if the set is too large, each noise vector appears in too few rows, making the noise-cluster anti-pattern nearly identical to noise anti-patterns. 
After carefully studying $12,000$ generated matrices with anti-patterns, we have observed that the set size approximately equals the average number of rows and columns across all visual patterns.
\jiangning{This appropriate set size ensures that noise-cluster anti-patterns remain distinct from both meaningful visual patterns and random noise.}

After generating the anti-patterns, we apply them sequentially to the matrix template.
For each entry affected by noise, its new value should be chosen from all possible values that differ from the original.
As a result, in binary matrices, the noise negates entries, while in continuous matrices, it replaces entries with random values.

\vspace{1.5mm}
\noindent\textbf{Swapping indices}.
To further increase the diversity of the matrix variations, we randomly swap their rows and columns to simulate real-world matrices that are not fully reordered.
The upper bound for the number of index swaps should be sufficient to make the order of matrix rows and columns completely random.
In \jiangning{Sec. 4 of the} supplemental material, we derive that this upper bound is $\frac{1}{2} n\log n$, where $n$ is the number of rows in the matrix.
As the number of index swaps increases, the rate of degeneration slows down, leading to a more gradual decline in the visual pattern quality.
Therefore, we use numbers that increase exponentially rather than linearly to ensure different degrees of quality in visual patterns.
Specifically, for each matrix variation, we apply varying numbers of index swaps, including $0$ and powers of 2 up to the closest one to $\frac{1}{2} n\log n$.

\subsection{Score Calculation}
\begin{figure}[!t]
\centering
{\includegraphics[width=\linewidth]{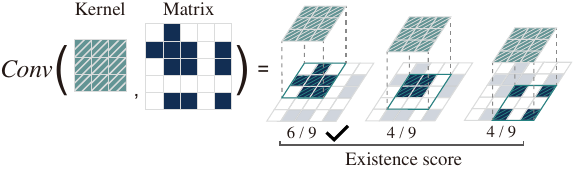}}
\vspace{-3mm}
\caption{The greedy matching strategy in the \jiangning{pattern-detection phase}. A convolutional kernel scans the matrix variation and the region with the highest \jiangning{existence} score is chosen.}
\vspace{-5mm}
\label{fig:conv}
\end{figure}

A straightforward solution to score the visual patterns in a matrix is to measure the entry-wise similarity with its template.
However, this method ignores the translation invariance of visual patterns.
For example, translating a block pattern along the main diagonal does not affect its visual quality.
However, it can significantly change the entry-wise similarity.
To tackle this issue, we consider how well the regions in the matrix can match a visual pattern and how well they present the visual pattern as a connected component.
Previous studies have shown that 
\begin{enumerate*}
\item convolutional kernels are capable of matching visual features in images~\cite{lecun1998gradient}, and 
\item entropy effectively quantifies the level of disorder in a set of potential outputs~\cite{bishop2006pattern}.
\end{enumerate*}
Building on these findings, we develop a convolution- and entropy-based scoring method that combines the matching capability of convolution and the disorder detection capability of entropy.
This scoring method consists of two phases: pattern detection and pattern scoring.
In the \jiangning{pattern-detection phase}, given a matrix variation and a visual pattern from its template, we utilize the matching capability of convolutional kernels to greedily match this pattern in the matrix variation and identify the region with the highest \jiangning{existence} score (Fig.~\ref{fig:conv}).
In the \jiangning{pattern-scoring phase}, we adjust this \jiangning{existence} score based on the connectedness of the matched region, an important aspect of its visual quality~\cite{palmer1994rethinking}.
The disorder detection capability of entropy is effective in quantifying this connectedness.
Therefore, we use it to calculate a disorder score for the matched region.
For continuous matrices, we further include the deviation score to measure the deviation of the matched region from the Robinson structure.
The final score of a matrix variation is derived from the scores of all matched regions.

\begin{figure}[!b]
\centering
\vspace{-3mm}
{\includegraphics[width=\linewidth]{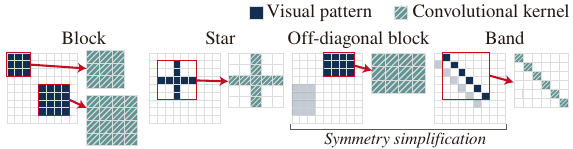}}
\vspace{-3mm}
\caption{The derivation of convolutional kernels from the matrix template.}
\label{fig:kernel}
\end{figure}

\vspace{1.5mm}
\noindent\textbf{Pattern detection}.
The \jiangning{pattern-detection phase} detects visual patterns in both binary and continuous matrices.
To facilitate this process, continuous matrices are converted to binary ones by setting non-zero entries to $1$\jiangning{, ensuring the focus remains on the existence of meaningful values rather than their exact magnitudes}.
Fig.~\ref{fig:kernel} \jiangning{illustrates the process of deriving convolutional kernels from each pattern in the matrix template.
A convolutional kernel is represented as a matrix containing only the pattern itself}.
In particular, for off-diagonal block and band patterns that consist of diagonal symmetric components, we simplify the kernel to contain only one of the components.
The convolutional kernel is then used to scan the matrix variation and calculate convolutions with its regions, as illustrated in Fig.~\ref{fig:conv}.
For block and star patterns, the kernel scans along the main diagonal.
For off-diagonal block and band patterns, the kernel scans the off-diagonal regions.
The region with the highest convolution that does not overlap with previously matched regions is chosen to match the pattern.
Since larger visual patterns reveal more meaningful information, the matching process starts with the largest patterns and proceeds to the smallest ones.

\vspace{1.5mm}
\noindent\textbf{Pattern scoring}.
In the \jiangning{pattern-scoring phase}, the \jiangning{existence} scores, disorder scores, and deviation scores are derived to measure the quality of visual patterns presented by the matched regions.
These scores are then aggregated to produce a final score for the matrix variation.

The \jiangning{existence} score (\jiangning{$S^e$}) is measured by the convolution between the matched region and the convolutional kernel.
To make the scores comparable across visual patterns, this convolution is normalized by the area of the matched region.
Formally, for a matched region with entries $\bm{a}=\{a_{i,j}\}$, the \jiangning{existence} score is calculated as $\jiangning{S^e}=\sum_{i,j} \mathbb{I}[a_{i,j}>0] /  |\bm{a}|$, where $\mathbb{I}$ is the indicator function that takes the value 1 if its argument is true and 0 otherwise, and $|\bm{a}|$ is the number of entries in the matched region, representing its area.
\jiangning{Rounding non-zero entries to 1 in this score ensures the focus remains on the existence of meaningful values rather than their exact magnitudes.}

\begin{figure}[!t]
\centering
{\includegraphics[width=\linewidth]{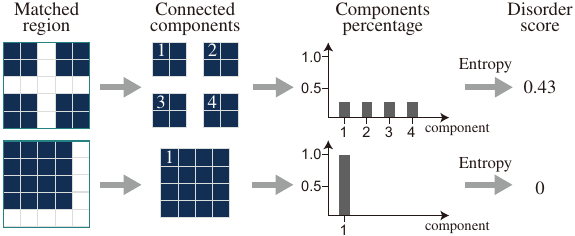}}
\vspace{-3mm}
\caption{The calculation of disorder scores based on entropy.}
\vspace{-5mm}
\label{fig:disorder}
\end{figure}

The disorder score ($S^d$) measures the degree to which the matched region is fragmented into smaller components by zero entries.
As shown in Fig.~\ref{fig:disorder}, this score is derived from the entropy of the proportions of the connected non-zero components.
\jiangning{In computing these components, we consider two entries connected if they share an edge or a corner.}
To make the scores comparable across visual patterns, we normalize the entropy by the logarithm of the area of the matched region.
Formally, for a matched region with $\ell$ connected components each occupying $pr_i$ percentage of the total non-zero entries, the disorder score is calculated as $S^d = (-\sum_{i=1}^\ell pr_i\cdot \log (pr_i))/ \log (|\bm{a}|)$.

\begin{figure}[!b]
\centering
\vspace{-3mm}
{\includegraphics[width=\linewidth]{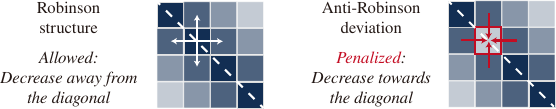}}
\vspace{-3mm}
\caption{In the Robinson structure, entries decrease away from the main diagonal. The deviation score measures the violation of this condition.}
\label{fig:continuous}
\end{figure}

The deviation score ($S^v$) \jiangning{is adapted from AR deviations~\cite{chen2002generalized} to evaluate all four types of visual patterns. 
It} quantifies the degree to which entries violate the Robinson condition of monotonically decreasing away from the diagonal (Fig.~\ref{fig:continuous}).
To make the scores comparable across patterns, we normalize the deviation score by the maximum possible deviation.
For block and star patterns, it is calculated as:

\vspace{-2mm}
\begin{equation}
\hspace{-2mm}
\label{eq:deviation_score}
\begin{aligned}
    S^v=\frac{\sum\limits_{i,j,k}f(a_{i,k},a_{i,j})\mathbb{I}[a_{i,k}<a_{i,j}]\mkern-2mu + \mkern-2mu f(a_{k,j},a_{i,j})\mathbb{I}[a_{k,j}<a_{i,j}]}{\sum\limits_{i,j,k} f(a_{i,k},a_{i,j})\mkern-2mu  + \mkern-2mu f(a_{k,j}.a_{i,j})}
\end{aligned}
\vspace{-2mm}
\end{equation}
Here, $i<k<j$. $\mathbb{I}[a_{i,k} < a_{i,j}]$ and $\mathbb{I}[a_{k,j} < a_{i,j}]$ indicate whether two entries violate the Robinson condition, while $f(a_{i,k}, a_{i,j})$ and $f(a_{k,j}, a_{i,j})$ measure the degree of violation.
Specifically, $f(a_{i,k}, a_{i,j})=\mathbb{I}[a_{i,k}>0]\mathbb{I}[a_{i,j}>0] |a_{i,k}-a_{i,j}|$ ensures that the violation is counted only when both entries are non-zero, as the disorder score already heavily penalizes the zero entries for fragmenting the components.
For off-diagonal block patterns\jiangning{, there are multiple diagonals along which the Robinson structure can be formed.
To address this ambiguity}, we identify the diagonal that minimizes the deviation score, \jiangning{which most accurately reflects the underlying Robinson structure~\cite{chen2002generalized}}.
Therefore, we calculate the score as the minimum of $S^v$ values derived from all diagonals.
Please refer to \jiangning{Sec. 5 of the} the supplemental material for more details.
For binary matrix variations, we set the deviation scores to 0.

\begin{figure}[!t]
\centering
{\includegraphics[width=\linewidth]{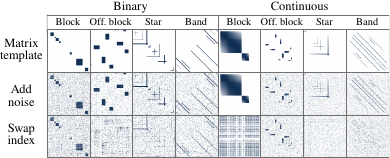}}
\vspace{-3mm}
\caption{Examples of matrix templates and variations in ReorderBench.}
\vspace{-5mm}
\label{fig:examples}
\end{figure}

A matched region that simultaneously achieves a high \jiangning{existence} score, a low disorder score, and a low deviation score presents a high-quality visual pattern.
Thus, the quality score ($S$) of a matched region is calculated as: 

\vspace{-2mm}
\begin{equation}
\label{eq:pattern_score}
\begin{aligned}
    S = \jiangning{S^e} \cdot (1-S^d) \cdot (1-S^v).
\end{aligned}
\vspace{-2mm}
\end{equation}
Larger visual patterns reveal more meaningful information.
Therefore, for a matrix variation with $p$ matched regions, the final quality score is weighted by their areas:

\vspace{-2mm}
\begin{equation}
\begin{aligned}
    \sum_{k=1}^p \frac{|\bm{a}_k|}{\sum_{l=1}^p |\bm{a}_l|} \cdot S_k,
\end{aligned}
\vspace{-2mm}
\end{equation}
where $S_k$ and $|\bm{a}_k|$ are the quality score and area of the $k$-th matched region, respectively.

\subsection{Statistics}
We have generated the largest benchmark, ReorderBench, for \jiangning{matrix-reordering} tasks.
It contains $8,505,000$ matrices, including $2,835,000$ binary and $5,670,000$ continuous matrices.
Each matrix comes with a score to reflect the quality of visual patterns. 
These matrices are of four sizes: [$100\PLH100$, $200\PLH200$, $300\PLH300$, $400\PLH400$].
The upper bound for the matrix size is set to $400\PLH400$ because it approaches the maximum size that can be effectively displayed on standard screens.
For instance, on a display with a resolution of 1920×1080, each entry in such matrices can be only $\lfloor 1080/400 \rfloor=2$ pixels wide.
Fig.~\ref{fig:examples} shows examples of generated matrices.
For more examples, please refer to \jiangning{Sec. 6 of the} the supplemental material and our website at \href{https://reorderbench.github.io/}{https://reorderbench.github.io/}.

\begin{figure}[!b]
\centering
\vspace{-3mm}
{\includegraphics[width=\linewidth]{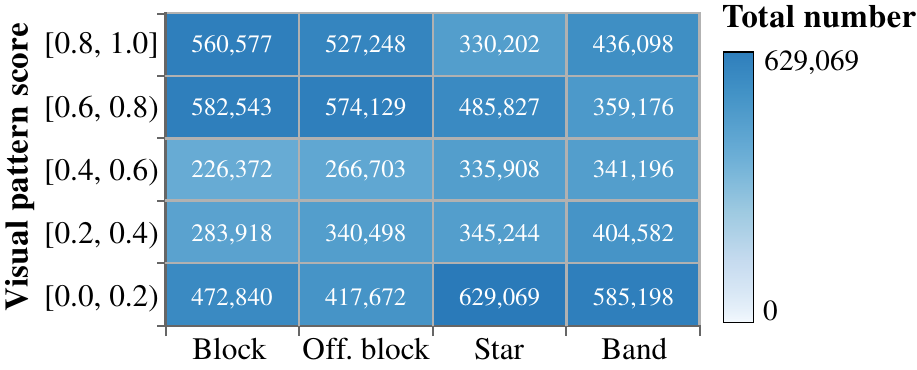}}
\vspace{-3mm}
\caption{Visual pattern types vs. Visual pattern scores.}
\label{fig:score_stat}
\end{figure}

In ReorderBench, the matrix distribution across different attributes, such as noise level and size, is relatively balanced.
For example, variations in the percentage of matrices at each noise level do not exceed $0.05\%$.
Fig.~\ref{fig:score_stat} shows that the generated matrices effectively cover different score ranges of visual patterns.
While star and band patterns tend to have more low-score matrices due to their susceptibility to noise and index swaps, the overall distribution of matrices across pattern types and scores remains balanced, with the largest difference being $629,069:226,372\approx 2.78:1$.
This is noticeably lower than the commonly referenced $4:1$ standard~\cite{krawczyk2016learning}.
\jiangning{Additionally, matrices in ReorderBench exhibit strong diversity. 
For example, 95$\%$ of randomly selected pairs of matrix templates with block patterns exhibit a Jaccard similarity~\cite{jaccard1901etude} of less than $0.31$.
The full results for all patterns are available in Fig. 3 of the supplemental material.}

\jiangning{To ensure consistent and reproducible evaluation on ReorderBench, we split it} into a training set of $6,804,000$ matrices and a test set of $1,701,000$ matrices while maintaining the same distribution across attributes in both sets.
To ensure that reordering methods are evaluated on matrix templates unseen during training, we assign each matrix variation to the same set as its corresponding template.

To accommodate tasks that require different distributions, adjustments can be made by sampling or generating more matrices for specific types.
\jiangning{For example, in Sec. 8 of the supplemental material, we demonstrate how our methods can be adapted to overlapping block patterns.}
\jiangning{To complement the generated matrices, we include $450$ real-world matrices from the Network Repository~\cite{rossi2015network}, the TUDataset~\cite{morris2020tudataset}, and the LRGB Dataset~\cite{dwivedi2022long} in ReorderBench and annotate their visual patterns.
Please refer to Sec. 9 of the supplemental material for more details.}
Overall, our benchmark offers a comprehensive resource catering to the needs of both researchers and practitioners.

%% file: 5-dataset-statistics.tex
\section{Data Study on Scoring Consistency}
\label{sec:study}
Consistency with human assessments of visual pattern quality greatly influences the usefulness of the scoring method.
Therefore, we evaluate how well our convolution- and entropy-based method align with expert assessments.
In alignment with prior research~\cite{panavas2023investigating}, we conduct an empirical data study where a few selected experts examine
many data, in contrast to the traditional method where a large group of ordinary people reviews limited data. 
We invite three experts, each with extensive experience in matrix reordering. 
The first expert is a senior researcher at a major IT company whose research interests include computer vision and visual analytics. 
He often uses matrix visualization to analyze the log data produced by machine learning models. 
The other two experts are fifth-year Ph.D. students majoring in visual analytics who are not co-authors of this paper.
They have used various reordering algorithms in their research projects. 
\jiangning{The second and third experts collaborate occasionally. 
There is no in-depth collaboration between the authors and any of the experts.}
\jiangning{We conduct the data study on both generated and real-world matrices.
The study settings and results for the generated matrices are presented here, while those for the real-world matrices are provided in Sec. 9.3 of the supplemental material.}

\subsection{Study Settings}
\noindent\textbf{Baseline}. We compare our scoring method against existing quality metrics,
which either \jiangning{measure the similarity between
adjacent entries in the matrix or focus on the distances between the
rows and columns in the ordering}~\cite{van2021simultaneous}.
Since there is no consensus on their comparative effectiveness, we select commonly used quality metrics for our analysis, \jiangning{including Moran's \textit{I}~\cite{van2021simultaneous}, LA~\cite{petit2003experiments}}, AR events~\cite{chen2002generalized} and BAR~\cite{earle2015advances}.

\noindent\textbf{Task and experiment design}.
To gather the assessments of human experts on visual pattern quality, the task is to compare different variations generated from a given matrix.
In the pilot study, for each visual pattern, we sample three binary matrices and three continuous matrices without index swaps, resulting in a total of 24 matrices.
For each matrix, we generate eight different variations and evaluate them using both baseline metrics and our method.
Four of the variations are generated by applying $0$, $8$, $64$, and $512$ index swaps, respectively.
The other four variations are generated by optimizing the four baseline metrics \jiangning{using a simulated annealing-based algorithm, ARSA~\cite{brusco2008heuristic}, which exclusively optimizes each metric to ensure the focus on comparing them}.
We create questions for all pairwise comparisons of the eight variations, resulting in 28 questions per matrix.
Thus, we have 
$
3(experts) \times 24(matrices) \times 28(questions) = 2,016
$
results from the pilot study.
In the formal study, we sample $48$ matrices evenly distributed across patterns, leading to $4,032$ results in the formal study.
The matrices involved in the data study are provided in \jiangning{Figs. 7, 8, and 9 of} the supplemental material.
Based on the gathered assessments, we compare the level of consistency between each quality metric and human experts.

\noindent\textbf{Study website and randomization protocol}.
The study is conducted through a web-based prototype that presents the questions across pages.
On each page, two variations to be compared are placed side by side.
The experts are then asked to assess which variation includes the higher-quality visual pattern (left, right, or indistinguishable).
To help the experts focus on judging one pattern at a time, questions are structured to form groups based on the pattern type and matrix type (binary or continuous).
To eliminate the learning effect, the questions in each group are presented in random order.

\subsection{Pilot Study}
This study aims to build a shared understanding of high-quality visual patterns among the experts, ensuring higher inter-rater reliability for the subsequent formal study. 
Following the common practice~\cite{panavas2023investigating, chang2017revolt, sun2014evoriver}, this study involves continuous discussions among the experts to refine and validate evaluation criteria.
Upon completing the pilot study, the expert assessments are gathered to verify the inter-rater reliability.

A widely used measure for inter-rater reliability is the intraclass correlation coefficient (ICC)~\cite{koo2016guideline}.
Following the guidelines provided by Koo~\etal~\cite{koo2016guideline}, we use the two-way mixed effects, absolute
agreement, single rater/measurement ICC.
The pilot study achieves an ICC of 0.814 with a 95\% confidence interval of [0.791, 0.834].
According to Landis~\etal~\cite{landis1977measurement}, this result indicates substantial agreement among the three experts.
This shared understanding of high-quality visual patterns serves as a strong foundation for the formal study.


\begin{figure}[!b]
\centering
\vspace{-3mm}
{\includegraphics[width=\linewidth]{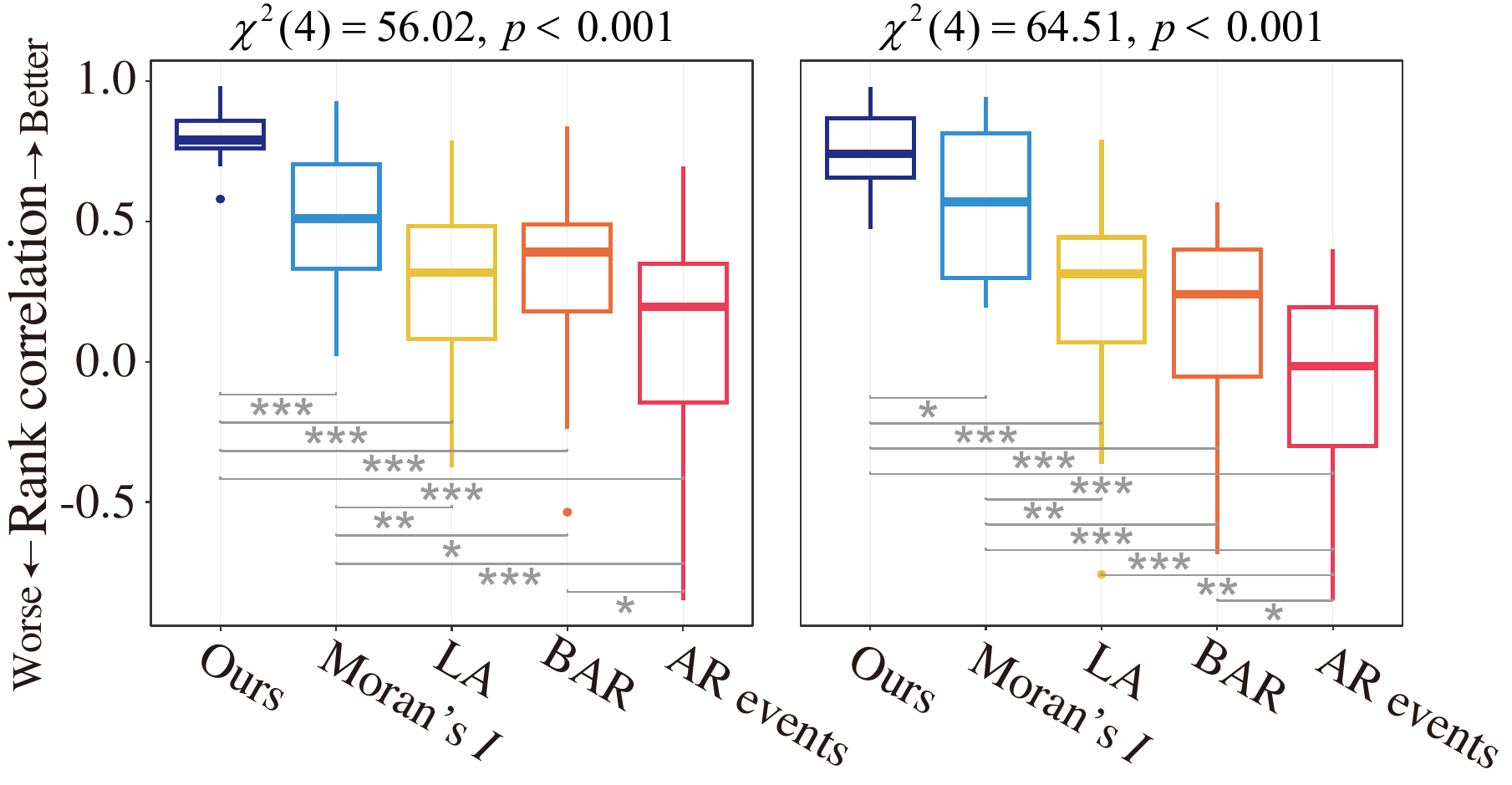}}
\vspace{-3mm}
\caption{Friedman tests and pairwise Wilcoxon signed-rank tests among the quality metrics. The left plot shows the results for binary matrices, while the right plot is for continuous matrices.
Asterisks indicate significance levels: {\fontfamily{phv}\selectfont
*}
indicates $p < 0.05$, {\fontfamily{phv}\selectfont
**} indicates $p < 0.01$ , and {\fontfamily{phv}\selectfont
***} indicates $p < 0.001$.}
\label{fig:study}
\end{figure}

\subsection{Formal Study}

In this study, the experts answer the questions independently.
They achieve an ICC of \jiangning{0.789} with a 95\% confidence interval of \jiangning{[0.772, 0.806]}.
This level of agreement ensures the reliability of expert assessments for further analysis.

To compare the quality metrics with expert assessments, we first rank the variations according to the comparison results of each matrix.
One issue in assigning the ranks is that, despite the substantial agreement among the experts, several conflicts still arise regarding the comparison results.
To address this, we employ the Elo rating system~\cite{elo1978rating} to handle such conflicts.
This system adjusts the ratings of the variations to align with the comparison results, thereby reflecting the quality judged by the experts.
The variation with the highest rating is ranked $1$, with subsequent ranks assigned in descending order of rating.
After assigning the ranks, we use Kendall's coefficient of rank correlation to measure the level of consistency between each quality metric and human experts.

Fig.~\ref{fig:study} shows the distribution of the rank correlation on binary and continuous matrices.
Following the common practice in empirical studies~\cite{zhou2023cluster, chen2024dynamic}, 
we perform Friedman tests to compare the rank correlation of the quality metrics.
The results show that the correlation difference among the five quality metrics is significant in both binary matrices ($\chi^2(4)=$ \jiangning{$56.02$}, $p<0.001$) and continuous matrices ($\chi^2(4)=$ \jiangning{$64.51$}, $p<0.001$).
The pairwise Wilcoxon signed-rank test further indicates that the convolution- and entropy-based scoring method significantly outperforms existing metrics ($p<0.05$).
These results demonstrate that our scoring method aligns with the experts in assessing the visual pattern quality.

%% file: 6-use-cases.tex
\section{Benchmark Applications and Analysis}
\label{sec:cases}

In this section, we illustrate the potential utility of our benchmark on three applications. 
First, it enables the evaluation of reordering algorithms by comparing their performance on the benchmark.
Second, it supports the development of a deep scoring model for measuring the quality of visual patterns. 
Third, it enables the development of a \jiangning{matrix-reordering} model based on the metric provided by the scoring model.
\jiangning{The evaluations presented here are conducted on the generated matrices. 
For those on the real-world matrices, please
refer to Sec.~9.4 of the supplemental material.}

\subsection{Evaluating Reordering Algorithms}
\label{subsec:benchmark}

We compare \jiangning{$45$} reordering algorithms on the ReorderBench test set to assess their effectiveness in revealing visual patterns.

\begin{table}[b]
\fontsize{8}{8}\selectfont
\renewcommand\arraystretch{1.1}
\setlength{\tabcolsep}{.2em}
\centering
\vspace{-4mm}
\caption{Evaluated base \jiangning{matrix-reordering} algorithms.}
\vspace{-2mm}%
\begin{tabular}{cc}
\toprule
Category & Algorithm \\ \midrule
\multirow{2}{*}{Robinsonian} & 
ARSA~\cite{brusco2008heuristic}, 
Dendser~\cite{earle2015advances}, GW~\cite{gruvaeus1972two}\\
& HC~\cite{sokal58statistical}, OLO~\cite{bar2001fast}, QAP~\cite{wilhelm1987solving} \\
Spectral & R2E~\cite{chen2002generalized}, Spectral~\cite{ding2004linearized} \\
Dim. reduction & 
LLE~\cite{roweis2000nonlinear}, MDS~\cite{kruskal1964multidimensional}, PCA~\cite{abdi2010principal}  \\
Heuristic & 
Barycenter~\cite{eades1994edge}, Moment~\cite{deutsch1971ordering}  \\
\multirow{2}{*}{Graph} &
\jiangning{NN$\_$2OPT~\cite{van2021simultaneous}},
OPTICS~\cite{ankerst1999optics}, 
RCM~\cite{cuthill1969reducing}, \\
& SPIN~\cite{tsafrir2005sorting},
TSP~\cite{hahsler2007tsp},   VAT~\cite{bezdek2002vat}\\
\bottomrule
\end{tabular}
\label{table:algorithms}
\end{table}

\vspace{1.5mm}
\noindent\textbf{Reordering algorithms}. Behrisch~\etal~\cite{behrisch2016matrix} classify existing automated reordering algorithms into 6 categories: Robinsonian, spectral, dimension reduction, heuristic, graph-theoretic, and biclustering.
We select commonly used reordering algorithms in each category except for biclustering, as they are not suitable for symmetric matrices.
This results in \jiangning{45} algorithms, including \jiangning{19} base algorithms and their variations.
The selected base algorithms are listed in Table~\ref{table:algorithms}.
The full list of evaluated algorithms is available in \jiangning{Table 2 of} the supplemental material.
The implementations of 39 algorithms are sourced from the R package seriation~\cite{hahsler2008getting}. 
We use the RCM algorithm from the Python SciPy library~\cite{virtanen2020scipy}, \jiangning{the NN$\_$2OPT algorithm from the Reorder.js library~\cite{fekete2015reorder}}, and implement 4 algorithms in the heuristic category in Python for which we cannot find any reference implementation.

\vspace{1.5mm}
\noindent\textbf{Evaluation criterion}. 
After obtaining the reordering results from all the evaluated algorithms, we compute their performance scores using the convolution- and entropy-based scoring method. 
In particular, the matrix without index swaps serves as the ground truth for its variations with index swaps. 
The performance score of a reordered matrix is defined as the ratio of its quality score to that of the ground-truth matrix. 
We then calculate the performance score of an algorithm by averaging those of all its reordered matrices.

\begin{table}[!t]
\fontsize{8}{8}\selectfont
\renewcommand\arraystretch{1.1}
\setlength{\tabcolsep}{.2em}
\centering
\caption{Evaluation results of existing \jiangning{matrix-reordering} algorithms and our \jiangning{matrix-reordering} model described in Sec.~\ref{subsec:deep_reorder}. The best one is \textbf{bold}, and the runner-up is \underline{underlined}.}
\vspace{-2mm}%
\begin{tabular}{ccccccccc}
\toprule
\multirow{2}{*}{Method}
& \multicolumn{2}{c}{Block} & \multicolumn{2}{c}{Off. Block} & \multicolumn{2}{c}{Star} & \multicolumn{2}{c}{Band}  \\
&Binary&Cont.&Binary&Cont.&Binary&Cont.&Binary&Cont. \\
\midrule
\multicolumn{9}{c}{\textbf{Robinsonian}} \\
GW\_ward~\cite{gruvaeus1972two} & 0.886 & 0.815 & 0.783 & 0.724 & 0.457 & 0.333 & 0.269 & 0.189\\
OLO\_ward~\cite{bar2001fast} & 0.892 & \underline{0.836} & \underline{0.795} & \underline{0.751} & 0.475 & 0.367 & 0.275 & 0.194\\
\multicolumn{9}{c}{\textbf{Spectral}} \\
R2E~\cite{chen2002generalized} & 0.830 & 0.628 & 0.661 & 0.559 & 0.445 & 0.380 & 0.245 & 0.176 \\
Spectral\_norm~\cite{ding2004linearized} & 0.726 & 0.517 & 0.556 & 0.443 & 0.344 & 0.237 & 0.228 & 0.161
 \\
\multicolumn{9}{c}{\textbf{Dimension reduction}} \\
MDS~\cite{kruskal1964multidimensional} & 0.763 & 0.539 & 0.623 & 0.525 & 0.441 & 0.368 & 0.238 & 0.171
 \\
PCA~\cite{abdi2010principal} & 0.763 & 0.539 & 0.623 & 0.525 & 0.441 & 0.368 & 0.238 & 0.171 \\
\multicolumn{9}{c}{\textbf{Heuristic}} \\
Barycenter~\cite{eades1994edge} & 0.460 & 0.360 & 0.477 & 0.414 & 0.400 & 0.264 & \underline{0.304} & \underline{0.252}
 \\
Moment~\cite{deutsch1971ordering} & 0.557 & 0.437 & 0.403 & 0.318 & 0.297 & 0.166 & 0.219 & 0.159
\\
\multicolumn{9}{c}{\textbf{Graph}} \\
BEA\_TSP~\cite{lenstra1975some} & 0.852 & 0.780 & 0.723 & 0.722 & 0.433 & 0.329 & 0.273 & 0.190
 \\
SPIN~\cite{tsafrir2005sorting} & \underline{0.895} & 0.727 & 0.766 & 0.613 & \underline{0.495} & \underline{0.427} & 0.247 & 0.167
 \\
\multicolumn{9}{c}{\textbf{Deep reordering model (ours)}} \\
Deep model (ours) & \textbf{0.925} & \textbf{0.926} & \textbf{0.861} & \textbf{0.912} & \textbf{0.828} & \textbf{0.854} & \textbf{0.757} & \textbf{0.759} \\
\bottomrule
\end{tabular}
\vspace{-6mm}
\label{table:methods_comparison}
\end{table}

\vspace{1.5mm}
\noindent\textbf{Results and analysis}.
We evaluate existing reordering algorithms on matrices of all four sizes.
To identify the most effective algorithms within each algorithm category, we apply a selection process based on the performance averaged over matrix sizes.
The selection favors algorithms with the best performance on individual patterns and those that rank as the top two in performance further averaged over all patterns.
In our case, this results in two selected algorithms per category since the top-performing algorithm on individual patterns overlaps with one of the top two on average performance across patterns.
Table~\ref{table:methods_comparison} shows the average performance of these algorithms. 
The full results are available in \jiangning{Sec. 11 of} the supplemental material.
Both the average results and individual results on each size demonstrate that existing algorithms perform well on block and off-diagonal block patterns but poorly on star and band patterns.
This points to the research potential for developing algorithms that can effectively reveal star and band patterns.
The two algorithms that achieve the best performance for block, off-diagonal block, and star patterns are OLO\_ward and SPIN.
OLO\_ward, a variation of the optimal leaf ordering algorithm~\cite{bar2001fast}, refines the hierarchical clustering dendrogram obtained with Ward's linkage~\cite{ward1963hierarchical}. 
Hierarchical clustering captures the global structure of the matrix. 
Meanwhile, the optimal leaf ordering algorithm optimizes the distance between adjacent leaves, which further performs local optimization to reveal visual patterns.
SPIN, the sorting points into neighborhoods algorithm~\cite{tsafrir2005sorting}, 
iteratively relocates each row to its most suitable neighborhood. 
Inspired by simulated annealing, the method starts by exploring large-scale relocations to build the global layout, then gradually reduces the relocation scale to capture the local structure.
The commonality of the two methods in considering both global and local structures highlights the importance of integrating these two perspectives to reveal visual patterns.
Although Barycenter achieves the best performance for band patterns, its average performance score ($0.366$) is considerably lower than OLO\_ward ($0.573$) and SPIN ($0.542$) and ranks sixth-to-last among the \jiangning{$45$} evaluated algorithms.
Therefore, we exclude it from our discussion.

\subsection{Building a Deep Scoring Model}
\label{subsec:measure}
\begin{figure}[!tb]
\centering
{\includegraphics[width=\linewidth]{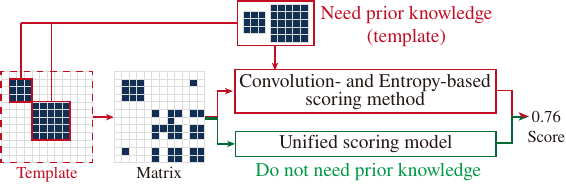}}
\vspace{-3mm}
\caption{Input of the scoring method and the unified scoring model.}
\vspace{-5mm}
\label{fig:scoring}
\end{figure}

The accuracy of our scoring method in measuring the quality of visual patterns makes it an appealing optimization criterion for reordering algorithms.
However, this method requires prior knowledge of the types and sizes of the patterns to configure the convolutional kernels.
This prerequisite, often absent in matrices beyond this benchmark, limits the generalizability of the scoring method.
To address this, we build a unified scoring model based on ReorderBench.
This model aligns with the convolution- and entropy-based scoring method across all four visual patterns in both binary and continuous matrices and can also measure matrices of varying sizes.
As shown in Fig.~\ref{fig:scoring}, the key feature of this model is that it does not require prior knowledge of the patterns in the matrix, which greatly improves its generalizability \jiangning{to matrices beyond ReorderBench}.
Next, we introduce how to build this model.

\vspace{1.5mm}
\noindent\textbf{Selecting candidate deep neural networks}. 
The shared utilization of the convolution mechanism makes a convolutional neural network (CNN) a promising choice to better align with the convolution- and entropy-based scoring method. 
Therefore, we narrow down the choice of deep neural networks to CNNs.
We test three commonly used CNNs:
ResNet-50~\cite{he2016deep}, VGG-16~\cite{simonyan2014very}, and ConvNeXt-T~\cite{liu2022convnet}.

\begin{table*}[!t]
\fontsize{8}{8}\selectfont
\renewcommand\arraystretch{1.1}
\setlength{\tabcolsep}{.3em}
\centering
\caption{The performance of deep neural networks as the unified scoring model. The best one is in \textbf{bold}.}
\vspace{-2mm}%
\begin{tabular}{c|cccccccc|c}
\toprule
\multirow{3}{*}{Model} & \multicolumn{8}{c|}{Scoring accuracy (mean absolute error)} & \multirow{3}{*}{Time (ms)} \\
&\multicolumn{2}{c}{Block} & \multicolumn{2}{c}{Off. Block} & \multicolumn{2}{c}{Star} & \multicolumn{2}{c|}{Band}&   \\
&Binary&Cont.&Binary&Cont.&Binary&Cont.&Binary&Cont.\\
\midrule
ResNet-50 & 0.0224 & 0.0160 & 0.0327 & 0.0238 & 0.0488 & 0.0388 & 0.0399 & 0.0557 & 3.446\\
VGG-16  & 0.0238 & 0.0170 & 0.0314 & 0.0233 & 0.0456 & 0.0357 & 0.0380 & 0.0520 & 5.115\\
ConvNeXt-T & \textbf{0.0203} & \textbf{0.0150} & \textbf{0.0259} & \textbf{0.0195} & \textbf{0.0398} & \textbf{0.0312} & \textbf{0.0332} & \textbf{0.0471} & \textbf{3.420}\\
\bottomrule
\end{tabular}
\vspace{-4mm}
\label{table:scorer}
\end{table*}

\vspace{1.5mm}
\noindent\textbf{Model training}. 
Each ReorderBench sample includes: 
\begin{enumerate*}
\item matrix $A_i$,
\item the type of pattern in its template $pt_i$,
\item the ground-truth quality score $S_i$, and
\item the number of index swaps applied to the matrix $sw_i$.
\end{enumerate*}
A key characteristic of matrices with high-quality patterns is that similar rows are adjacent.
Existing scoring methods check this adjacency by taking the dissimilarity matrix as input~\cite{hahsler2008getting}.
Consequently, we augment each sample with dissimilarity matrix $D_i$. 
\jiangning{In line with common practice in matrix reordering~\cite{earle2015advances,behrisch2016matrix}, we derive the dissimilarity matrix $D_i$} from $A_i$ using the Euclidean distance between its rows.
\jiangning{This method is chosen for its simplicity, computational efficiency, and proven effectiveness in capturing the differences between rows in a matrix.}

The unified scoring model is trained to: 
\begin{enumerate*}
\item align with the convolution- and entropy-based scoring method for pattern type in the matrix, and
\item predict minimal scores for pattern types absent from the matrix.
\end{enumerate*}
To achieve the two objectives, the scoring model minimizes: 
\vspace{-2mm}
\begin{equation}
\hspace{-2mm}
\begin{aligned}
    L = \sum_{i=1}^N \sum_{j=1}^4 (\mathbb{I}[j = pt_i](\hat{S}_{i,j} - S_i)^2 + \mathbb{I}[j\neq pt_i] \mathbb{I}[sw_i = 0]\hat{S}_{i,j}^2),
\end{aligned}
\vspace{-2mm}
\end{equation}
where the first term is the alignment cost, and the second term is the absence penalty. $\hat{S}_{i,j}$ is the quality score predicted by the scoring model for the $j$-th pattern type of the $i$-th sample.

In the first term, to encourage the alignment of the scoring model with the convolution- and entropy-based scoring method, we minimize the squared difference between their respective quality scores.

In the second term, to penalize the scoring model for predicting non-zero scores for absent pattern types, a straightforward method is to assume all types of patterns other than the one in the template are absent.
However, the index swaps could introduce new types of patterns.
For instance, dividing the rows of a block pattern into two consecutive segments yields two separate block patterns and two off-diagonal block patterns.
Consequently, we exclude matrices with index swaps from the penalization term.

We fine-tune our unified scoring model from pre-trained models on ImageNet-1k~\cite{deng2009imagenet}. 
We train the model for 10 epochs with a batch size of 512 using AdamW optimizer~\cite{loshchilov2018decoupled}.
The initial learning rate is set to 0.0001, and we employ a cosine annealing scheduler.

\vspace{1.5mm}
\noindent\textbf{Scoring matrices of varying sizes}. 
To predict scores for matrices of varying sizes, the key is transforming the matrix to the desired size while preserving visual patterns.
Previous research in computer vision has shown that resizing an image effectively achieves this goal and incorporates it as a routine step in image pre-processing~\cite{radford2021learning}.
Therefore, to score matrices of different sizes, we convert them to images and resize them to the training size of the model.
\jiangning{We resize the matrices using OpenCV~\cite{opencv_library} with the INTER$\_$AREA interpolation option, as this option best preserves the visual patterns based on our detailed examination of $700$ resized matrices.}
The score is then predicted by the model based on the resized images. 
During training and evaluation, we resize all matrices to $200\PLH 200$ \jiangning{as an example to demonstrate the effectiveness of our unified scoring model. 
We have verified that for $100$ larger matrices from the Network Repository with sizes up to $2000\times 2000$, downsampling to $200\times 200$ effectively preserves major visual patterns in all of them. 
For tasks involving even larger matrices, a new scoring model with a higher input resolution can be easily trained. 
For results on training scoring models with higher input resolutions, please refer to Sec. 12 of the supplemental material.}

\vspace{1.5mm}
\noindent\textbf{Results}. 
Based on the three CNNs, we build three scoring models and compare their accuracy and efficiency in scoring.
The scoring accuracy is measured by the mean absolute error between the quality score from the model and the convolution- and entropy-based scoring method. 
The scoring efficiency is measured by the time required to score a matrix using an NVIDIA Geforce RTX 3060 GPU.
In addition, we demonstrate the capability of the model to handle real-world matrices of varying sizes. 

\begin{figure}[!t]
\centering
{\includegraphics[width=\linewidth]{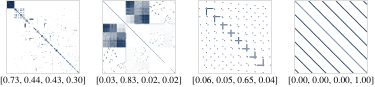}}
\vspace{-3mm}
\caption{Quality scores predicted by our scoring model, which measure the quality of block, off-diagonal block, star, and band patterns, respectively.}
\vspace{-5mm}
\label{fig:scoring_model}
\end{figure}

We evaluate the three scoring models on matrices of all four sizes.
Table~\ref{table:scorer} shows their average performance.
The full results are available in \jiangning{Sec. 11 of} the supplemental material.
On all matrix sizes, ConvNeXt-T consistently performs the best in terms of both accuracy and efficiency.
Therefore, we employ ConvNeXt-T to build the unified scoring model.
We demonstrate the capability of the scoring model to handle real-world matrices by applying it to the Pajek graph collection~\cite{batagelj1998pajek} and the Network Repository~\cite{rossi2015network}. 
To align with the training samples, we normalize these matrices so that their entries fall within $[0,1]$.
As shown in Fig.~\ref{fig:scoring_model}, the unified scoring model generates reasonable scores. 
\jiangning{For example, the first matrix contains a high-quality block pattern with a score of $0.73$. 
This score represents the percentage of the identified block pattern relative to the inherent ones in this matrix.}

\subsection{Building a \jiangning{Matrix-Reordering} Model}
\label{subsec:deep_reorder}

The representative and diverse matrices in ReorderBench offer valuable supervision for training deep reordering models. 
\jiangning{Existing deep learning-based reordering methods are typically trained on limited datasets that consist of either the given matrix~\cite{watanabe2022deep} or its reordering results~\cite{kwon2022deep}. 
These training methods aim to learn features specific to the given matrix, resulting in models that are adept at reordering only the given matrix and thus lack generalizability. To address this limitation, we} treat the \jiangning{index-swapped matrices} as negative samples and their \jiangning{corresponding} ground-truth matrices as positive samples. \jiangning{This strategy allows our model to learn a broader range of features, facilitating the ability to generalize across previously unseen matrices.} 
\jiangning{The model is based }on ResNet~\cite{he2016deep} due to its demonstrated performance. 
The model architecture is introduced in \jiangning{Sec. 13 of} the supplemental material.

\vspace{1.5mm}
\noindent\textbf{Model training}. The ReorderBench training set is denoted by $\{(A_i, D_i, \tilde{A}_i)\}_{i=1}^N$, where for the $i$-th sample, $A_i$ is the matrix, $D_i$ is the dissimilarity matrix, and $\tilde{A}_i$ is the ground-truth matrix.
The model aims to reconstruct $\tilde{A}_i$ by reordering $A_i$.
Since the matrices take value in $[0,1]$, we use the binary cross-entropy loss as the reconstruction loss:
\vspace{-2mm}
\begin{equation}
\hspace{-2mm}
\begin{aligned}
    L = -\sum_{i=1}^N \sum_{j,k} (\tilde{a}_i)_{j,k} log(\hat{a}_i)_{j,k} + (1-{(\tilde{a}_i})_{j,k}) log(1-(\hat{a}_i)_{j,k}),
\end{aligned}
\vspace{-2mm}
\end{equation}
where $(\tilde{a}_i)_{j,k}$ are entries of the $i$-th ground-truth matrix and $(\hat{a}_i)_{j,k}$ are entries of the reordered matrix of the $i$-th matrix. 

We train eight deep reordering models for each matrix size, with four dedicated to binary matrices and four to continuous matrices. 
Within each matrix type, four models correspond to four different visual patterns.
All these models are trained with a batch size of 512 using AdamW optimizer with a base learning rate of 0.01.
For each model, we train for 120 epochs and employ a cosine annealing scheduler with 20 epochs of linear warm-up.
\jiangning{The models trained on each matrix type are integrated into an ensemble-based method~\cite{breiman1996bagging}, where the unified scoring model selects the best reordering result from these models.
We provide more details on the ensemble-based method in Sec. 9.4 of the supplemental material.}
To further enhance the model performance and robustness, we perform test-time augmentation based on the unified scoring model.
More details on the augmentation are described in \jiangning{Sec. 14 of} the supplemental material.

\vspace{1.5mm}
\noindent\textbf{Results}.
The last row of Table~\ref{table:methods_comparison} presents the performance of our deep reordering model averaged over all matrix sizes. 
The full results are available in \jiangning{Sec. 11 of} the supplemental material.
For the block pattern, our model performs slightly better than existing methods.
As a central topic for \jiangning{matrix-reordering} research, the task of revealing block patterns is well studied and poses a challenge for further improvement. 
For the off-diagonal block, star, and band patterns, our deep reordering model achieves higher performance scores than the existing methods.
Although ResNet-18 may seem basic by current standards, the observed improvement confirms the value of our benchmark.
Moreover, although reordering continuous matrices is generally harder than binary matrices, our model achieves higher performance scores for all four visual patterns in continuous matrices due to the larger amount of training data.
This also highlights the importance of large-scale datasets for training high-performance deep reordering models.

%% file: 7-discussion.tex
\section{Discussion and Future Work}
\label{sec:discussion}
As evidenced by our experimental findings, the main advantage of ReorderBench lies in its ability to facilitate comprehensive comparisons of reordering performance among different algorithms. 
Another advantage is that it facilitates the development of a unified scoring model, which inspires the development of new reordering methods.
Moreover, the representative and diverse matrices and their associated quality scores provide a test base for designing deep learning models that improve matrix reordering. 

Despite its benefits, ReorderBench still has several limitations that serve as starting points for future research on enhancing both the benchmark and the reordering techniques.

\begin{figure}[!b]
\centering
\vspace{-3mm}
{\includegraphics[width=\linewidth]{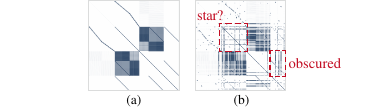}}
\vspace{-3mm}
\caption{Issues with hybrid patterns: (a) the initial matrix; (b) after index swaps, the patterns obscure each other, and star patterns appear.}
\label{fig:hybrid}
\end{figure}

\vspace{1.5mm}
\noindent\textbf{Support hybrid patterns}. 
Although the proposed pipeline achieves impressive results in generating matrices with the four visual patterns, the creation of matrices with hybrid patterns, such as the combination of block + band, poses two challenges.
The first is how to generate representative matrices with hybrid patterns. 
Addressing this involves exploring methods to integrate different patterns into hybrid ones, which deserves further investigation.
A promising method is to adapt the mix-up technique~\cite{zhang2018mixup}, commonly used to augment training data. 
This technique improves the robustness and generalization of the model by blending multiple images and their labels to create new composite samples. 
The second lies in assessing hybrid patterns.
As different types of visual patterns in a matrix can potentially obscure each other, and new patterns may appear during variation generation, it is difficult to measure the quality of these hybrid patterns by simply combining their scores derived by our scoring method.
For example, Fig. ~\ref{fig:hybrid}(a) shows a matrix with a hybrid pattern of off-diagonal block and band.  
After 30 times of index swaps, these two visual patterns obscure each other, and many star patterns also appear (Fig. ~\ref{fig:hybrid}(b)).
This highlights the importance of exploring effective methods to identify and measure the quality of hybrid patterns when they are mixed during the variation process.
\jiangning{Although our initial attempts to handle hybrid patterns, presented in Sec. 9 of the supplemental material, are promising, fully addressing these challenges will require significant further effort and deserves a separate publication.}

\vspace{1.5mm}
\noindent\textbf{Support non-symmetric matrices}.
ReorderBench focuses on generating symmetric matrices and evaluating their visual patterns.  
Meanwhile, non-symmetric matrices are also involved in several fields, such as directed graphs in network analysis.
Therefore, exploring methods for their generation and evaluating the quality of their visual patterns is beneficial.
Extending the benchmark to include non-symmetric matrices would necessitate the development of specialized processing techniques. 
These techniques would ensure that the templates are properly generated and then degenerated into variations.
Furthermore, it is necessary to develop a more generalized scoring method that can effectively evaluate the visual patterns in non-symmetric matrices. 
This might include adapting the scoring method or inventing new ones specifically designed for the unique characteristics of these non-symmetric matrices, such as patterns that represent directional relationships.

\vspace{1.5mm}
\noindent\textbf{Reorder matrices of varying sizes with one model}.
Our evaluation demonstrates that, compared with existing algorithms, deep models greatly improve reordering performance by predicting the best permutation instead of performing a step-by-step search.
However, the adaptability of deep reordering models to matrices with varying sizes remains an issue. 
These models are generally designed for and trained on matrices of fixed size~\cite{kwon2022deep,watanabe2022deep}.
This highlights a potential area for innovation, developing foundation models to accommodate matrices of different sizes~\cite{yang2023foundation}.
For smaller matrices, padding the matrix with empty rows and columns allows it to be reordered by the deep model.
However, trials on our reordering model show that it occasionally ignores the matrix with smaller patterns, as the addition of many empty rows and columns makes these patterns nearly indistinguishable from noise. 
Thus, developing techniques to preserve original patterns during padding warrants further exploration.
For larger matrices, one possible solution is to employ the divide-and-conquer strategy. 
The matrix is first divided into smaller matrices, each of which is reordered independently. 
Then they are combined into the final reordered matrix. 
However, how to divide the matrix without damaging the visual patterns remains an issue, which deserves further exploration.

%% file: 8-conclusion.tex
\section{Conclusion}
\label{sec:conclusion}
In this paper, we introduce ReorderBench, a benchmark designed to evaluate and improve \jiangning{matrix-reordering} techniques.
It features a representative and diverse collection of binary and continuous matrices, and incorporates a convolution- and entropy-based scoring method. 
Three applications have demonstrated that this benchmark not only serves as a resource for evaluating different reordering algorithms but can also enable more effective and efficient development of a unified scoring model and deep reordering models.
These improvements enhance our ability to reveal and interpret the underlying structures within complex datasets.